\documentclass[aps,prd,twocolumn,amsfonts,amssymb,amsmath,showpacs,letterpaper,superscriptaddress,nofootinbib,altaffilletter,fontenc]{revtex4-1}

\usepackage[citecolor=blue,linkcolor=blue,colorlinks=true,urlcolor=blue]{hyperref}

\usepackage{color}
\usepackage{graphicx}
\usepackage{latexsym}
\usepackage{float}
\usepackage{amsmath}
\usepackage{amssymb}
\usepackage{multirow}
\usepackage{ifthen}
\usepackage{slashed}
\usepackage{lineno}
\usepackage{makecell} 
\usepackage[mathscr]{euscript} 
\usepackage{mathrsfs} 
\usepackage{subfigure}
\usepackage{stackrel}

\usepackage{longtable}
\usepackage{dcolumn}
\usepackage{array}
\usepackage{ragged2e}
\usepackage{multirow}
\usepackage[utf8]{inputenc}

\usepackage{tabularx,booktabs}
\newcolumntype{Y}{>{\centering\arraybackslash}X}
\usepackage{booktabs}
\usepackage{pifont}

\usepackage{array}

\newcolumntype{R}[1]{>{\RaggedRight}p{#1}}

\usepackage[normalem]{ulem}

\renewcommand{\today}{\number\day\space\ifcase\month\or
  January\or February\or March\or April\or May\or June\or
  July\or August\or September\or October\or November\or December\fi
  \space\number\year}

\def\be{\begin{equation}}
\def\ee{\end{equation}}
\def\bi{\begin{itemize}} 
\def\ei{\end{itemize}}
\def\ben{\begin{enumerate}}
\def\een{\end{enumerate}}

\def\prd{Phys. Rev. D.}

\begin{document}

\title{Using supervised learning algorithms as a follow-up method in the search of gravitational waves from core-collapse supernovae}

\author{Javier~M.~Antelis}
\email[E-mail: ]{mauricio.antelis@ligo.org}
\affiliation{Embry-Riddle Aeronautical University, Prescott, AZ 86301, USA}
\affiliation{Tecnologico de Monterrey, Escuela de Ingeniería y Ciencias, Monterrey, N.L., 64849, México}

\author{Marco~Cavaglia}
\affiliation{Institute of Multi-messenger Astrophysics and Cosmology, Missouri University of Science and Technology, Rolla, MO 65409, USA}

\author{Travis~Hansen}
\affiliation{Embry-Riddle Aeronautical University, Prescott, AZ 86301, USA}

\author{Manuel~D.~Morales}
\affiliation{Universidad de Guadalajara, Guadalajara, Jal., 44430, M\'exico}

\author{Claudia~Moreno}
\affiliation{Embry-Riddle Aeronautical University, Prescott, AZ 86301, USA}
\affiliation{Universidad de Guadalajara, Guadalajara, Jal., 44430, M\'exico}

\author{Soma~Mukherjee}
\affiliation{The University of Texas Rio Grande Valley, Brownsville, TX 78520, USA}

\author{Marek~J.~Szczepa\'nczyk}
\affiliation{University of Florida, Gainesville, FL 32611, USA}

\author{Michele~Zanolin}
\affiliation{Embry-Riddle Aeronautical University, Prescott, AZ 86301, USA}

\begin{abstract}
We present a follow-up method based on supervised machine learning (ML) to improve the performance in the search of gravitational wave (GW) burts from core-collapse supernovae (CCSNe) using the coherent WaveBurst (cWB) pipeline. The ML model discriminates noise from signal events using as features a set of reconstruction parameters provided by cWB. Detected noise events are discarded yielding to a reduction of the false alarm rate (FAR) and of the false alarm probability (FAP) thus enhancing of the statistical significance. We tested the proposed method using strain data from the first half of the third observing run of advanced LIGO, and CCSNe GW signals extracted from 3D simulations. The ML model is learned using a dataset of noise and signal events, and then it is used to identify and discard noise events in cWB analyses. Noise and signal reduction levels were examined in single detector networks (L1 and H1) and two detector network (L1H1). The FAR was reduced by a factor of $\sim10$ to $\sim100$, there was an enhancement in the statistical significance of $\sim1$ to $\sim2\sigma$, while there was no impact in detection efficiencies.
%
%
\end{abstract}


\maketitle

\section{Introduction}
\label{sec:Introduction}

%
The search and characterization of gravitational wave (GW) bursts with the network of laser interferometers LIGO~\cite{TheLIGOScientific:2014jea}, VIRGO~\cite{TheVirgo:2014hva} and KAGRA~\cite{Aso:2013eba} need to address issues such as discrimination between GW events and noise artifacts, reconstruction of the GW waveforms, and localization of the source in the sky.
In the case of GWs generated by binary black holes (BBH) and binary neutron stars (BNS), the existing algorithms benefits from having highly deterministic signal models, and thus, searches are based on match-filtering detector strain data with available template signals (see for example~\cite{LIGOScientific:2018mvr,Abbott:2020niy}).

%
Core-collapse supernovae (CCSNe) are also a primary detection target in the upcoming LIGO and VIRGO observing runs. 
%
%
CCSNe are of special interest because the electromagnetic radiation and emission of neutrinos along with GWs will provide new hints to understand their formation mechanism and dynamic, and also will lead to novel insights in multimessenger astronomy. 
%
%
%
%
%
The morphology of GWs from CCSNe is predominantly stochastic with some deterministic components like the growing trend of the central frequency of the fundamental g-f mode or the characteristic waveform during the core bounce emission in rapidly rotating progenitors.
Furthermore, the production of templates from three-dimensional (3D) state-of-the-art numerical simulations is computational extremely expensive, including the fact that multiple run simulations for same progenitor might produce slightly different morphologies, even if this limitation does not apply to the core bounce component of rapidly rotating progenitors that could be described by deterministic template banks~\cite{PhysRevD.95.063019}. 
These limitations are the main reasons why unmodeled or weakly modeled searches are so far the only approaches considered for CCSNe GWs~\cite{Abbott:2019pxc,PhysRevD.104.102002}.

%
%
Coherent WaveBurst (cWB) is a standard method for the search and characterization of GWs using data collected by the LIGO and VIRGO detectors without imposing assumptions about the signal morphology~\cite{DRAGO2021100678}.
To detect unmodelled GW transients, cWB identifies coincident excess-power between the strain data in the network of detectors using wavelet-based analyses~\cite{Necula_2012}.
Such coincidence between multiple detectors allows to identify and reject noise events since they are present in only one of the detectors.
The method also reconstructs the waveform of the detected GW and estimates some signal parameters~\cite{Klimenko2005,Klimenko2008,Klimenko2016}.
cWB has been an essential computational tool in the detection of many GW transients from binary systems discovered so far~\cite{Abbott:2016blz,TheLIGOScientific:2016uux,Abbott:2020tfl,Szczepanczyk:2020osv,Abbott:2020khf},
%
%
and has been a core method in all-sky and targeted searches of burst GW signals in the previous LIGO and VIRGO observing runs~\cite{Abbott:2015vir,Abbott:2017muc,Abbott:2016ezn,Abbott:2019prv,Abbott:2019pxc}.
Furthermore, cWB will play a critical role in the detection of CCSNe GWs in the upcoming fourth and fifth observing runs (O4 and O5) with the LIGO, VIRGO, and KAGRA (LVK) network~\cite{PhysRevD.104.102002}.

%
%
%
%
The search of GWs emitted by CCSNe possesses other difficulties in addition to the inherent uncertainties in the waveform models.
The production of candidate events is mostly driven by excess energy instances that can be related to stationary or non-stationary noise components.
For small signal-to-noise ratio (SNR) the stationary ones produce most of the events, while non-stationary ones (usually termed non-Gaussian glitches) dominate the large SNR values.
Morphologically both type of noise events tend to carry specific signatures of the physical causes that generated them.
The rate of more rare glitches has been largely reduced (see for example ref.~\cite{Abbott:2019pxc}) by requiring a temporal coincidence and some morphological similarity between different interferometers (as encoded in the so-called $cc$ coefficient of cWB).
However, if we have only one interferometer collecting data, temporal coincidence and morphological consistency are not available.
But even in the case of a network of detectors, the population of non-Gaussian glitches can have a rate sufficiently large to be a limiting factor in obtaining a large statistical confidence in the detection.

The events formed by cWB at small SNR tend to show a fairly compact time frequency structure that would be independent of the sources, however for increasing amplitudes the morphology can become more varied. 
The detector characterization group of the LVK network operate routinely to classify the morphology of different glitches~\cite{LIGODetectorCharacterization} and some of them, like the so-called blip glitches can strongly resemble the core bounce waveform for rapidly rotating progenitors~\cite{PhysRevD.104.102002}.
Even if cWB makes use of internal morphological tests with multiple interferometers (especially for extragalactic source locations), all the opportunities to improve either the detection confidence, the detection range or both need to be taken.
These issues encourage to identify more morphological metrics to distinguish CCSNe events from noise events, especially for loud glitches that can reduce the statistical confidence of a detection even when they are not temporally overlapped to the detection candidate.
Machine learning (ML) methods offers one of these opportunities and this paper is part of systematic the explorations of its potential.

%
In recent times, ML along with the special class of deep learning (DL) models have attracted a lot of attention to tackle several problems in the realm of GWs~\cite{Cuoco2020}, for instance, to discriminate between noise and GW signals either from binary systems~\cite{PhysRevLett.120.141103,PhysRevD.97.044039,Morales2020,PhysRevD.94.124040,PhysRevD.101.104003} or from CCSNe~\cite{Astone2018,PhysRevD.102.043022,Iess_2020,PhysRevD.103.063011}, and to identify and remove transient noise events using strain data or auxiliary channels ~\cite{Powell_2017,PhysRevD.88.062003,Razzano_2018,PhysRevD.101.042003}.
Notably, ML models also have been used to enhance cWB performance, in specific, to distinguish between glitches and GW signals from BBH~\cite{PhysRevD.102.104023}, to construct a statistical veto based on the recognition of noise events to improve the detection efficiencies of GWs from BBH~\cite{Mishra2020}, and to achieve higher detection sensitivity of GW signals from CCSNe using signal enhancement~\cite{Mukherjee:2017,PhysRevD.103.103008}.
ML models, in specific genetic programming algorithms, had been previously investigated to discriminate CCSNe GW signals from noise transients for the case of single detector searches~\cite{Cavaglia2020}.
This is significant because it has been shown that there are periods of time during the observing runs for which only one detector in the network is in operational conditions and collecting science-quality data~\cite{Cavaglia2020}. Indeed, $\sim$30\% of the collected strain data during the first observing runs has been from one detector only~\cite{Vallisneri_2015}. 
Along this line, this paper investigates the benefits of supervised ML as a follow-up method of cWB that discriminates between noise and signal events in searches of GWs from CCSNe.
We extended the previous approach~\cite{Cavaglia2020} by 
investigating the noise reduction in two detector network and assessing its effects in the detection range for extragalactic CCSNe;
%
%
using classification algorithms that rely on a discriminant function that do not require stochastic initialization of parameters and iterative executions thus avoiding to perform several training repetitions or the selection of a sub-optimal solution, as in the case of classifiers based on genetic approaches.

%
To discriminate between noise and signal triggers, we used as features a set of cWB reconstruction parameters such as duration and central frequency, and the supervised classification models linear discriminant analysis (LDA) and support vector machines (SVM).
%
%
Detected noise events are discarded, thus reducing the rate of false detections.
We carried out systematic studies to assess to quantify the effective improvement in the false alarm rate (FAR) and in the statistical significance with networks of one or two detectors using strain data from the O3a observing run, and with several recent CCSNe GW waveforms from 3D simulations.  
These studies considered robustness tests where the ML is tuned with a distribution of noise and signal triggers extracted from a stretch of data and then it is applied and tested in a different stretch of data.
%
%
%
The manuscript is organized as follows. 
Section~\ref{sec:Methods} describes the cWB analyses carried out to generate distributions of noise and signals triggers and the LDA and SVM classification models used to recognize between noise and CCSNe GW signal events.
Section~\ref{sec:results} presents the results of the studies devoted, first to assess the noise reduction and signal lost rates produced by the classification models, and second to ascertain the actual improvement in the FAR and the statistical significance.
Section~\ref{sec:conclusions} presents conclusions and future directions.

\section{Methods}
\label{sec:Methods}

\subsection{cWB analysis}
\begin{table*}
\caption{
Total time and the background (non-zero lag) time of each network of detectors (L1, H1 and L1H1) for the three stretches of open O3a LIGO data.}
\begin{tabular*}{\textwidth} {@{\extracolsep{\textwidth minus \textwidth}} c|cccccc} 
\hline
\hline
\multicolumn{1}{c|}{Time window}
& \multicolumn{2}{c}{TW1}
& \multicolumn{2}{c}{TW2}
& \multicolumn{2}{c}{TW3}
\\
\hline
\hline
\multicolumn{1}{c|}{Initial time}
& \multicolumn{2}{c}{2019-06-17T00:00:01}
& \multicolumn{2}{c}{2019-06-25T00:00:01}
& \multicolumn{2}{c}{2019-08-26T00:00:01}
\\
\multicolumn{1}{c|}{Final time}
& \multicolumn{2}{c}{2019-06-19T23:59:59}
& \multicolumn{2}{c}{2019-06-27T23:59:59}
& \multicolumn{2}{c}{2019-08-28T23:59:59}
\\
\hline
\hline
Network & Total time & Background time & Total time & Background time & Total time & Background time 
\\
L1      & 1.73 days & -            & 1.62 days & -            & 2.18 days & -            \\
H1      & 1.77 days & -            & 1.66 days & -            & 2.38 days & -            \\
L1H1    & 1.50 days & 4.6 years    & 1.06 days & 3.3 years    & 1.96 days & 6.2 years    \\
\hline
\hline
\end{tabular*}
\label{Table:CCSNe_TimeWindows}
\end{table*}
In this study we used LIGO strain data from the first half of the third observing run (O3a). 
cWB analyses were carried out independently in three stretches of open data (i.e., three time windows named TW1, TW2 and TW3) with a duration of three days each.
For the case of CCSNe GWs, the search is typically driven by optical observations where the time window ranges from hours to days, while it is a few seconds in the potential case of neutrino flux driven searches.
Hence, three days was chosen as a representative window duration of how much we would be able to constrain the GW emission from electromagnetic observations and a relatively rapid discovery of the CCSNe.
All cWB analyses were carried out separatelly with one and two detectors network (L1, H1, and L1H1) with the aim of studying the rate of false detection (background analysis), the detectability of CCSNe GWs (sensitivity analysis), and to generate datasets of noise and signal events to train and to test the classification algorithms.
Table~\ref{Table:CCSNe_TimeWindows} shows the total time of the three time windows for each network of detectors used in this study.
%

\subsubsection{cWB Background}
%
%
With more that one detector, the data from one detector is shifted with respect to the other in a time length that has to be longer than the GW travel time between detectors ($\sim$10ms between L1 and H1).
This time-shift procedure is repeated multiple times to obtain a total background search time (also called non-zero lag time) long enough to attach a certain statistical significance.
All detected events from this analysis are of non-astrophysical origin and therefore correspond to false detections or noise events.
Then, the FAR can be estimated simply as the ratio between the number of detected events and the total background time.
%
%
%
Table~\ref{Table:CCSNe_TimeWindows} also shows the total background time of the three time windows considered in this study.
Note that with only one detector available in the network the time-shifting is not possible, neither coincident test to remove glitches can be performed. This drastically worsen the FAR in comparison with networks of two or more operational detectors.

In the cWB background analysis, the FAR is produced for different values of the network SNR or $\rho$, yielding to the FAR versus $\rho$ curves.
Note that for the case of a single interferometer, $\rho$ is equivalent to the SNR of the available detector. 
%
%
The statistical significance is computed in terms of the false alarm probability (FAP) as follows~\cite{Abbott:2019pxc}:
\begin{equation}
    \label{eq:FAP}
	FAP = 1 - e^{T_{on} \times FAR},
\end{equation}
where $T_{on}$ is the on-source window where a GW signal is searched for.
%

\begin{figure*}[t]
    \centering
    \begin{tabular}{ccc}
    \includegraphics[width=0.32 \textwidth]{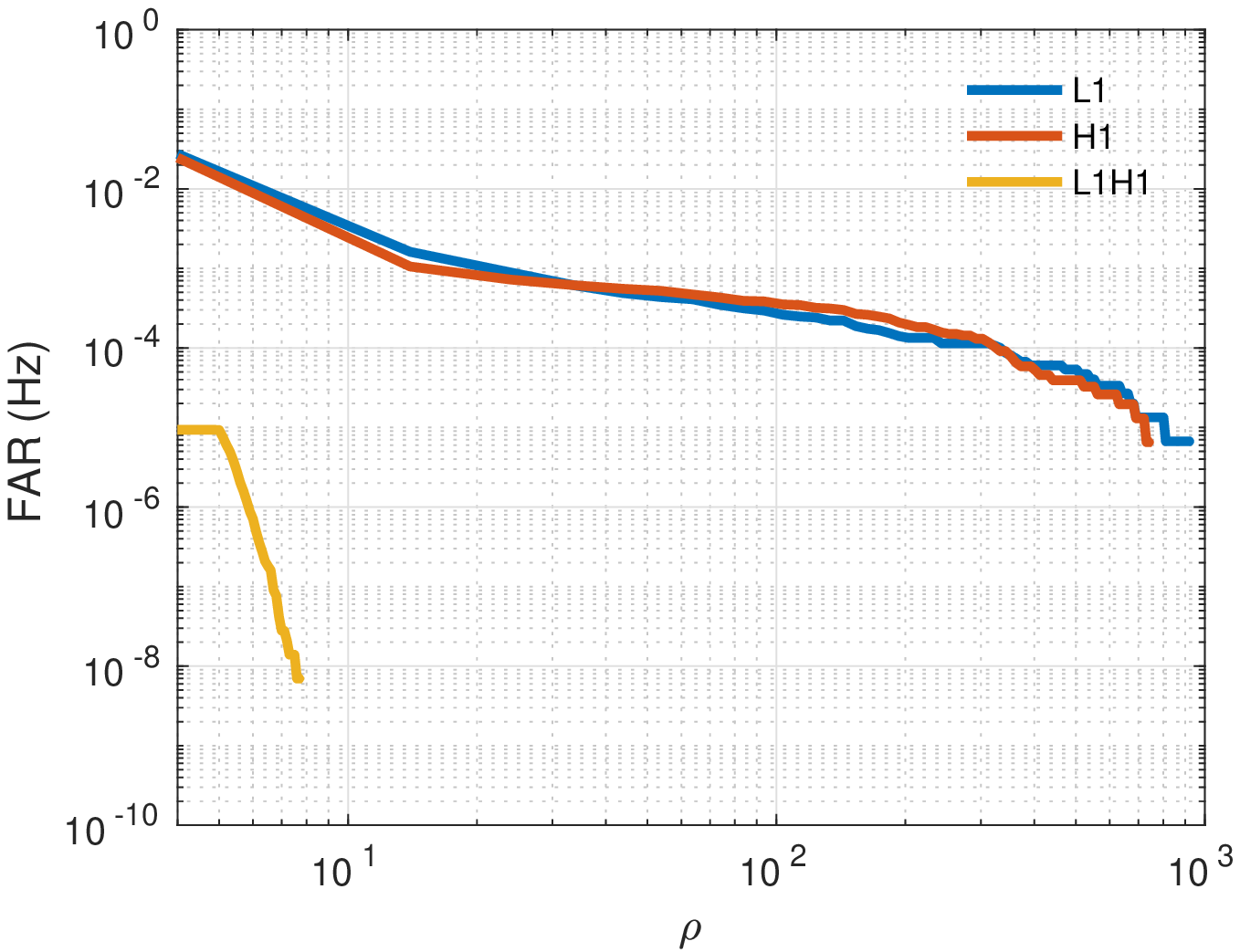}
    & 
    \includegraphics[width=0.32 \textwidth]{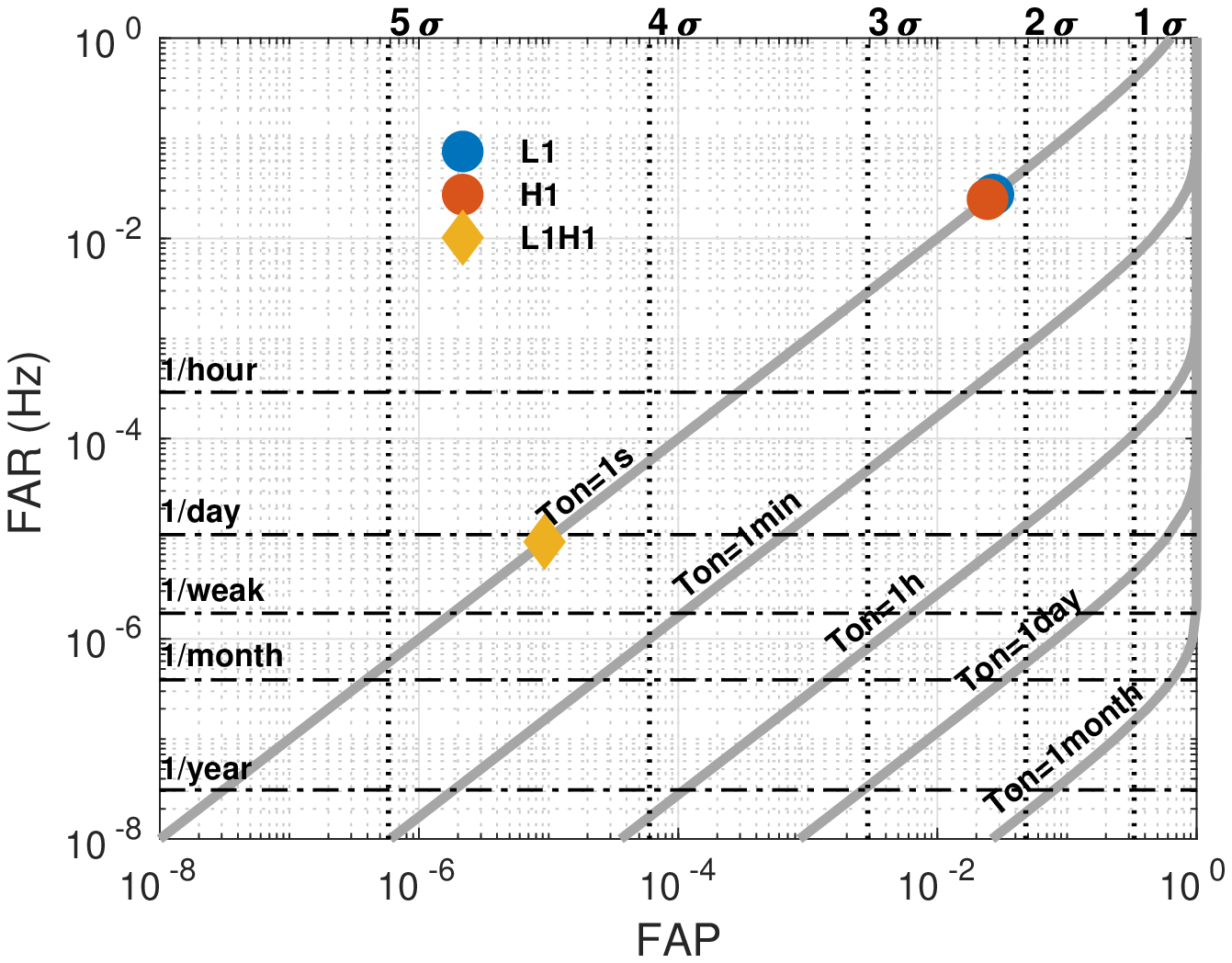}
    & 
    \includegraphics[width=0.30 \textwidth]{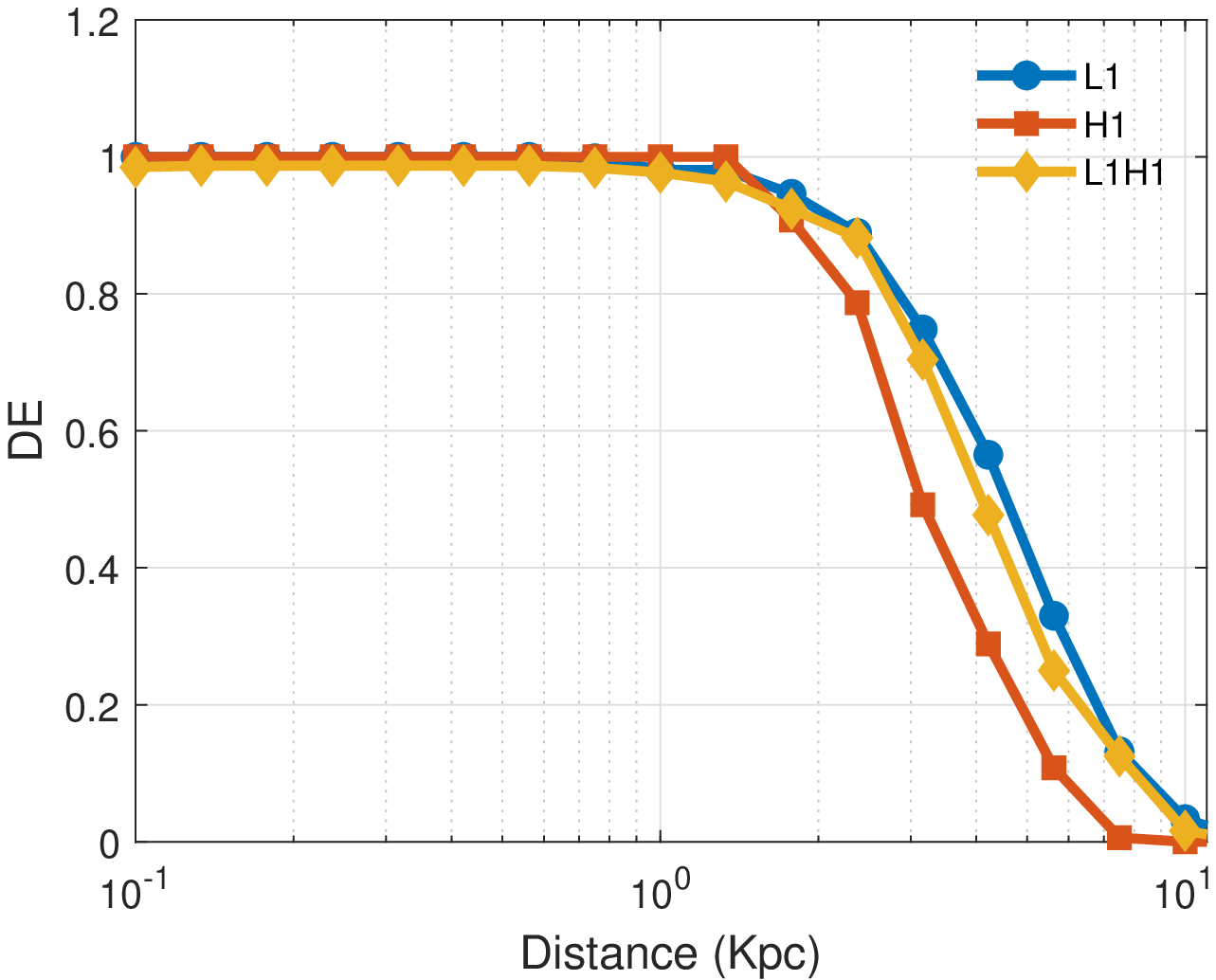}
    \\
    (a) & (b) & (c)
    \end{tabular}
	\caption{
	cWB search results for the cases of single detector network (L1 and H1) and two detectors network (L1H1) using strain data in the time window TW1.
    (a) False Alarm Rate (FAR) versus the $\rho$ statistic from the background analysis.
	(b) False alarm rate (FAR) versus False Alarm Probability (FAP) for the case of a on-source window of 1 second.
	(c) Detection efficiency (DE) with respect to the source distance achieved with the CCSNe GW ``he3.5" from \textit{Powell et al. 2019} for a corresponding FAR of $2.74\times10^{-2}$, $2.47\times10^{-2}$ and $9.30\times10^{-6}$ Hz for L1, H1, and L1H1, respectively.
	}
	\label{fig:cwbresults}
\end{figure*}

Results for the background analysis in the time window TW1 are presented in figure~\ref{fig:cwbresults}.
Note how the FAR (figure~\ref{fig:cwbresults}a) is considerably higher with one detector than with two detectors.
%
%
Likewise, the corresponding values of the FAP (figure~\ref{fig:cwbresults}b, where the $\rho$ threshold was set to 5) for the case of a search in a on-source window of 1s reveals a statistical significance barely close to $3\sigma$, and a quite higher significance close to $5\sigma$ for the two detector network. 
%
%
This illustrates the need of a follow-up method to identify and discard noise events (not only in single detector based searches but also in searches with two detectors) which reduces the FAR and improves the statistical significance of the search.

\subsubsection{cWB sensitivity}
The goal in this analysis is to determine the sensitivity in the search of known CCSNe GWs, and to obtain a distribution of signal events.
Here, waveforms are systematically added into detector noise data, rescaled to different amplitudes corresponding to diverse source distances, and at different time delays between detectors corresponding to diverse source sky locations.
For the case of single detector network, the injection is carried out only in the available detector and thus no time delay is required.
Subsequently, the search of GWs is carried out and the detection efficiency (DE) is simply measured as the fraction of successfully detected GWs.
DE is computed for each injected type of GW and for each distance to construct DE versus source distance curves.

All simulation analyses were done at different source distances from 0.1 up to 10kpc. 
%
%
Furthermore, several families of CCSNe GWs were used (see next subsection).
Results for the analysis around time window TW1 are presented in figure~\ref{fig:cwbresults}c for the case of the waveform named ``he3.5" from~\cite{Powell2019}.
These DE curves are for a FAR of $2.74\times10^{-2}$, $2.47\times10^{-2}$ and $9.30\times10^{-6}$ Hz for L1, H1, and L1H1, respectively.
%
This result shows how the efficiency reduces as the distance increases, and quite similar performance for all network of detectors.
Despite the good and steady sensitivity irrespective of the number of detectors, the low statistical significance imply not only to recognize the noise events with high accuracy, but also to not to affect these detection efficiencies.


\subsubsection{CCSNe GW waveforms}
To carry out simulation analyses, freely-available CCSNe GWs from recent 3D simulations were selected to be added into noise data.
We selected recent families of CCSNe GW waveforms that were used in the all-sky search of short GW bursts in O3~\cite{Abbott:2021Allskysearch}, and that were part of the recent study devoted to investigate the detectability of GWs from CCSNe in the upcoming fourth and fifth observing runs~\cite{PhysRevD.104.102002}.
In specific, we used the following waveform families representing a wide variety of physical phenomena and modeling methods:
\begin{itemize}
    \item \textit{Scheidegger et al. 2010}. This a large set of GWs obtained in 3D-simulations of magnetohydrodynamic (MHD) driven explosions with diverse rotating progenitors~\cite{Scheidegger2010}. For this work we considered three representative GWs generated by a progenitor start of 15$M_{\odot}$ with different rotational speeds and with neutrino leakage scheme, R1E1CA\_L (slowly rotating), R3E1AC\_L (moderate rotating), and R4E1FC\_L (rapidly rotating).
    \item \textit{O'Connor et al. 2018}. This is a family of seven GWs generated by a 20$M_{\odot}$ progenitor star mass~\cite{OConnor2018}. The simulations considered neutrino physics and the resulting GW signatures exhibit strong g-modes and SASI components. The name of the seven waveforms are mesa20, mesa20\_LR, mesa20\_pert, mesa20\_pert\_LR, mesa20\_v\_LR, mesa20\_2D, and mesa20\_2D\_pert. 
    \item \textit{Powell et al. 2019}. This is set of two GWs (named he3.5 and sl8) that were computed from simulations considering low to regular energies in the explosion mechanisms~\cite{Powell2019}. They are non-rotating models with 3.5 helium core solar masses after the star has lost its outer layers due to binary interactions for model he3.5 and with a a zero age main sequence (ZAMS) of 18$M_{\odot}$ for model sl8. The GW waveforms exhibit the typical frequency rise associated with emission of g-modes.
    %
    \item \textit{Powell et al. 2020}. This family of three GWs were obtained from simulations including explosion properties of the progenitor star~\cite{Powell2020}. The name of the waveforms are m39, s18np, y20. The first model is from a rapidly rotating progenitor of 39$M_{\odot}$, while the other two are from non-rotating progenitors of 18 and 20$M_{\odot}$. The model s18np has the same progenitor as the model s18 in Powell et al. 2019, but it does not include perturbations (np).
    %
    \item \textit{Mezzacappa et al. 2020}. This is a single GW named C15-3D generated by a 15$M_{\odot}$ progenitor start that includes a neutrino-driven convection mechanisms \cite{Mezzacappa:2020lsn}. The GW signature presents low- and high-frequency components and SASI emission.
\end{itemize}


\subsection{Features}
The features used to feed the classification algorithms to be able to discriminate between noise and signal events are reconstruction parameters provided by cWB for each event.
We used the same set of parameters proposed in~\cite{Cavaglia2020} which are representative to describe the duration, frequency and time-frequency characteristics of the reconstructed GW burst transients:
\begin{itemize}
    \item \textit{$\rho$}: cWB detection statistic.
    \item \textit{Volume}: number of time-frequency pixels composing the event.
    \item \textit{Duration 1} and \textit{Duration 2}: time length of the event computed from the energy-weighted and from the reconstructed waveform.
    \item \textit{Frequency 0} and \textit{Frequency 1}: central frequency of the event computed from the energy-weighted and from the reconstructed waveform.
    \item \textit{Low} and \textit{High}: minimum and maximum frequency of the time-frequency map pixels.
    \item \textit{Bandwidth 1} and \textit{Bandwidth 2}: bandwidth estimated from the energy-weighted and from the time-frequency map.
    \item \textit{Norm}: effective number of time-frequency resolutions used for GW reconstruction.
\end{itemize}
%
%
Hence, the vector of features is $\textbf{x} \in \mathbb{R}^{N_f \times 1}$ where $N_f$ is the number of reconstruction parameters extracted from each event.
%

\subsection{Classification algorithms}
The classifier is a computational model that takes as input a vector of features extracted from a cWB event, and assigns to it one class label indicating ``noise" or ``signal".
Common and robust classification models used in diverse applications are Linear Discriminant Analysis (LDA) and Support Vector Machines (SVM).
These classifiers allow to identify a linear or non-linear separation hyper-surface in such a way that the class assigned to a input vector of features depends on which region the vector is located~\cite{Duda2001}.
They consist of a discriminant function defined by $\textbf{w}^{T} \cdot f(\textbf{x}) = 0$, where $\textbf{x} \in \mathbb{R}^{N_f \times 1}$ is the vector of $N_f$ features, $f(\cdot)$ is a transformation function, and $\textbf{w} \in \mathbb{R}^{N_f \times 1}$ is a vector of classification weights (i.e., discriminant vector) that have to be calculated from a training dataset $\{\textbf{x}_i,y_i\}, i=1,...,N$, where $y_{i}$ indicates whether $\textbf{x}_i$ is a feature vector extracted from a noise or a signal event, and $N$ in the number of instances.

LDA defines the transformation function as $f(\textbf{x})=\textbf{x}$, thus it is only able to construct linear separation surfaces since the discriminant function becomes $\textbf{w}^{T} \cdot \textbf{x} = 0$~\cite{Bishop2006} .
In this classifier, the discriminant vector $\textbf{w}$ is estimated by seeking the projection that maximizes the difference between the means of the classes while minimizes their variance (leading to a classification model that is optimal when the two classes are Gaussian with equal co-variance).
Thus, the hyper-surface is found by solving this optimization problem:
\begin{equation}
    \label{eq:LDA1}
	\widehat{\textbf{w}} = \stackbin[\textbf{w}]{}{arg\,max} \frac{\textbf{w}^T\,\textbf{S}_B\,\textbf{w}} {\textbf{w}^T\,\textbf{S}_W\,\textbf{w}},
\end{equation}
where $\textbf{S}_B$ is the between-class co-variance matrix and $\textbf{S}_W$ is the within-class co-variance matrix~\cite{Bishop2006}. 
%
%
%

SVM is able to compute linear or non-linear separation surfaces in such a way that maximizes the separation between the hyper-surface and the nearest data points of each class which are called support vectors~\cite{Cortes1995}.
This involves solving the following optimization problem:
\begin{equation}
    \label{eq:SVM1}
	\widehat{\textbf{w}} = \stackbin[\textbf{w},\xi]{}{arg\,min}  \frac{1}{2}\|\textbf{w}\|^2 + C\sum_{i=1}^n\xi_i,
\end{equation}
subject to the condition $y_i \left( \textbf{w}^T f(\textbf{x}_{i}) \right) \geq 1 - \xi_i,  \forall i=1, ...,N$, where $\xi_i \geq 0$ measure the error in the misclassification of $\textbf{x}_{i}$.
The separation hyper-plane is linear when $f(\textbf{x}) = \textbf{x}$, leading to the linear SVM (or SVML). 
On the contrary, non-linear separation boundaries are obtained through a kernel function. The most common kernel is the radial basis function (RBF), which leads to a non-linear SVM with RBF (or SVMR), 

In this work, we first used LDA, SVML and SVMR to test the discrimination between noise and signal events, and then, we selected the one with the higher performance in order to assess the actual improvement produced by classification model as a follow-up method in offline cWB searches.

\section{Results}
\label{sec:results}

\begin{figure*}[t]
    \centering
    \begin{tabular}{ccc}
    \includegraphics[width=0.40 \textwidth]{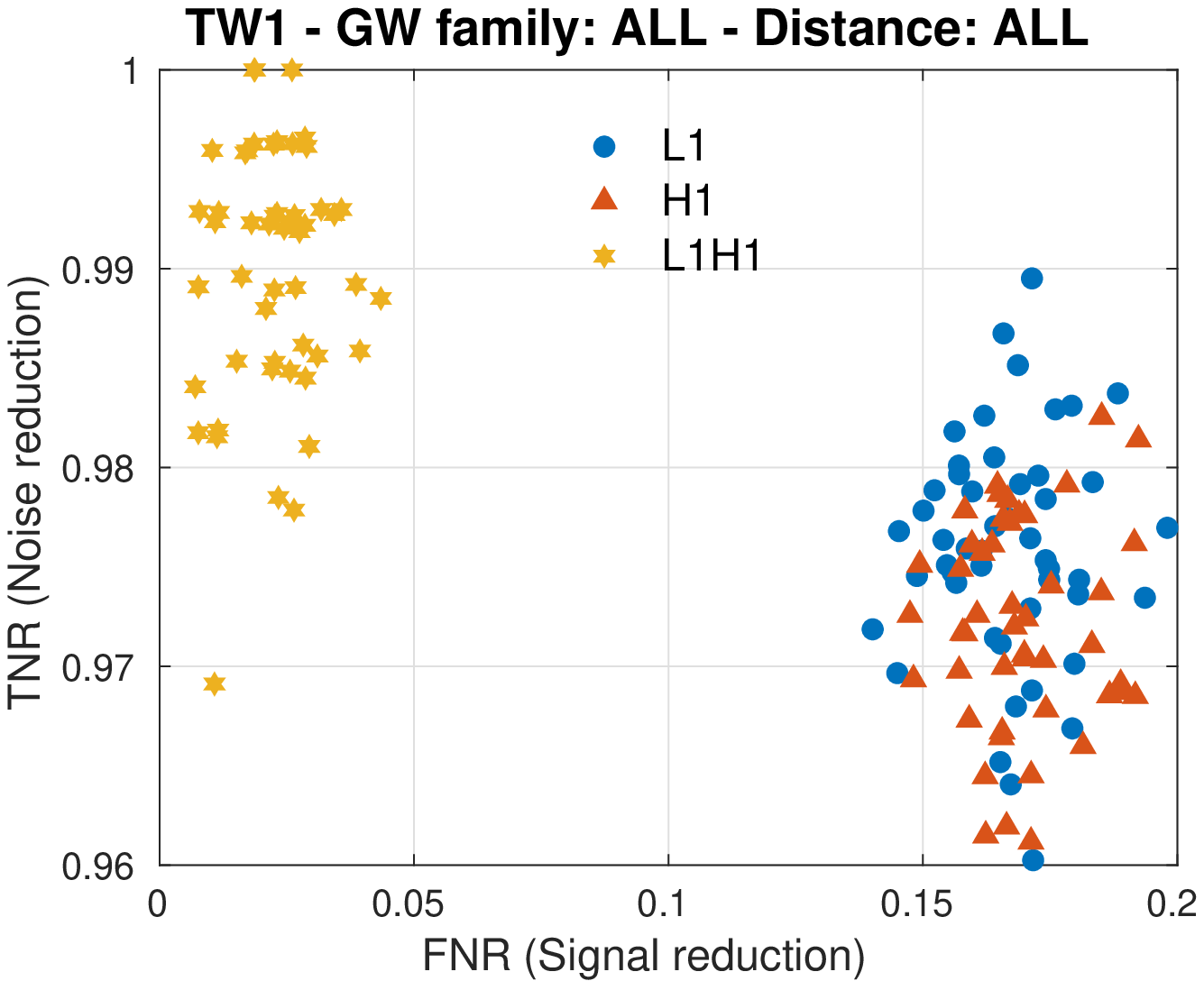}
    &
    \includegraphics[width=0.38 \textwidth]{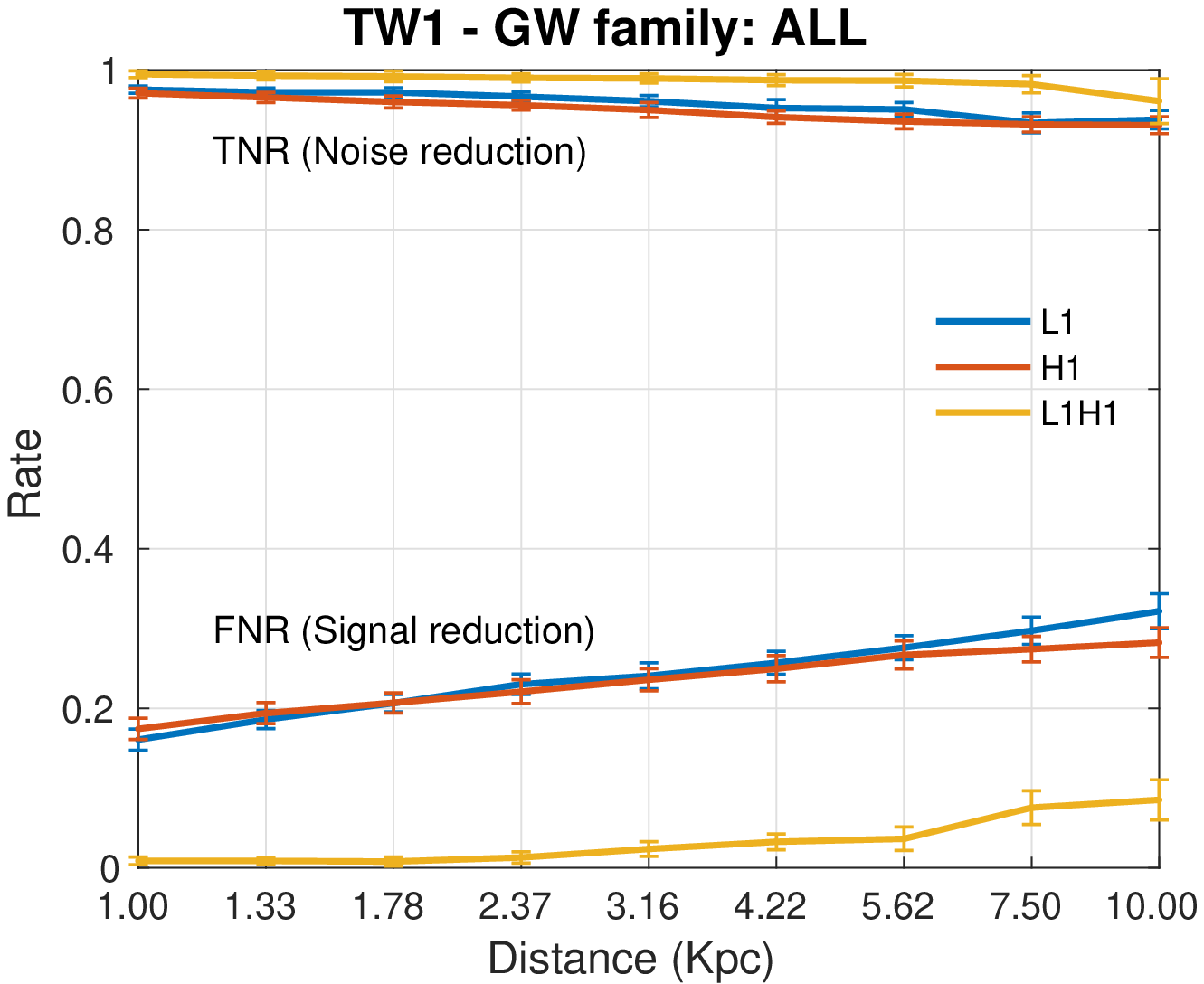}
    \\
    (a) & (b) 
    \end{tabular}
    \caption{
    Classification results from the cross-validation analysis obtained with the dataset of triggers extracted from the time window TW1 in detector networks H1, L1, and L1H1.
    (a) Distribution of classification rates (TNR and FNR) for the case of signal triggers from all families of GW and from all distances combined.
    (b) Mean and standard deviation values of classification rates (TNR and FNR) for the case of signal triggers from all families of GW and for each distance individually.
    All these results were obtained when using the classification algorithm SVMR.
    }
    \label{fig:CV_TNRandFNR}
\end{figure*}

\begin{figure*}[t]
    \centering
    \begin{tabular}{ccc}
    \includegraphics[width=0.28 \textwidth]{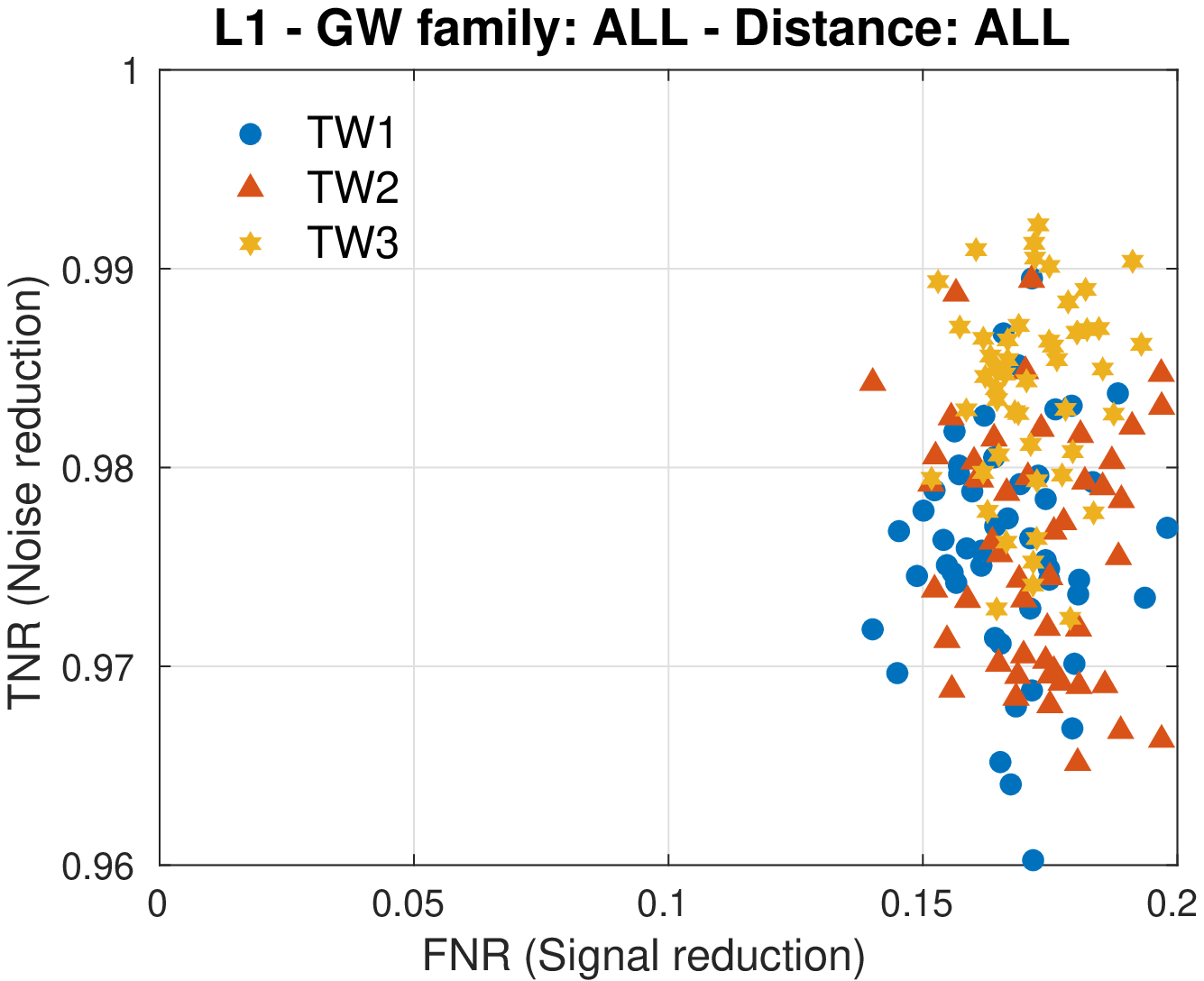}
    & 
    \includegraphics[width=0.28 \textwidth]{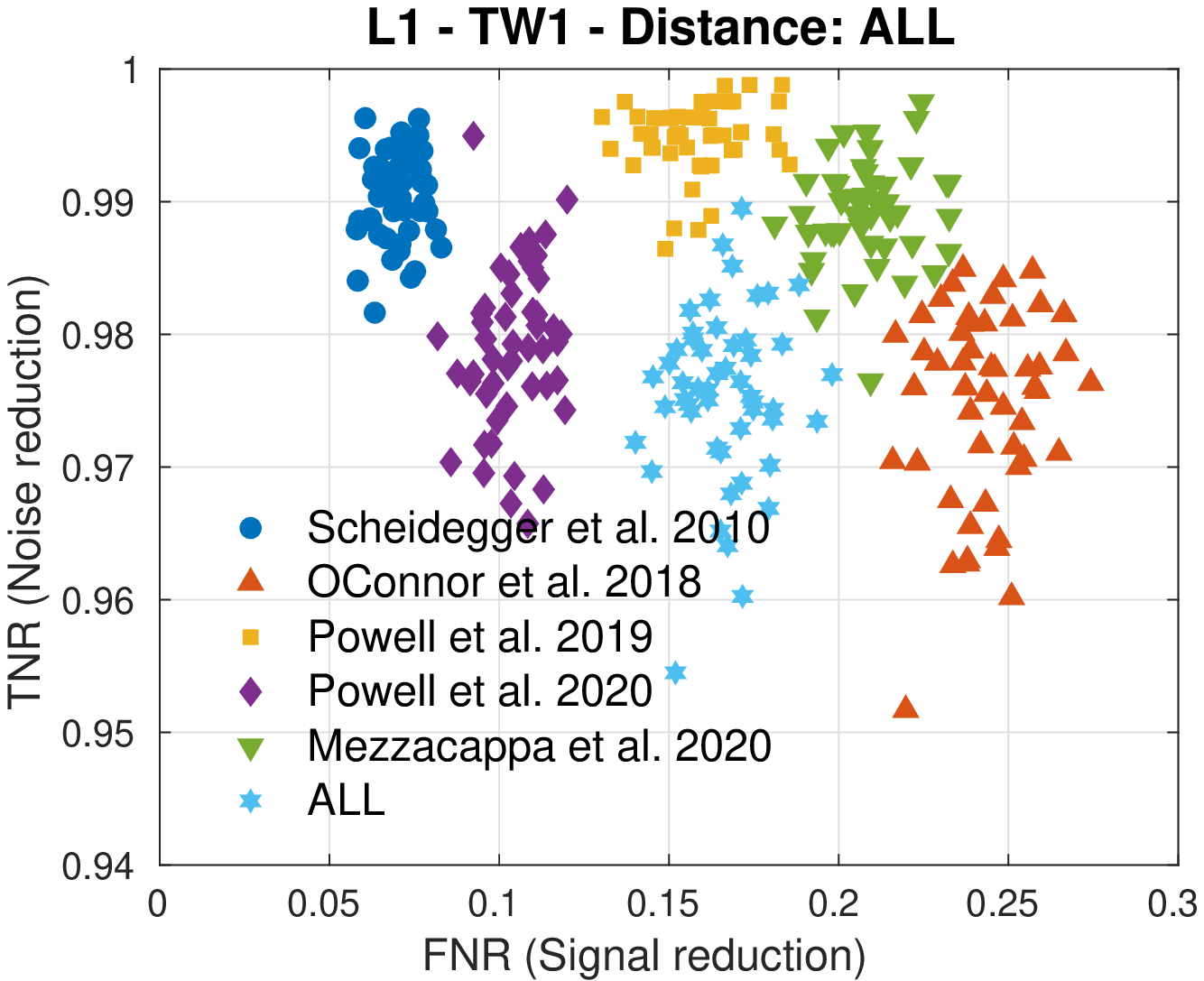}
    & 
    \includegraphics[width=0.305 \textwidth]{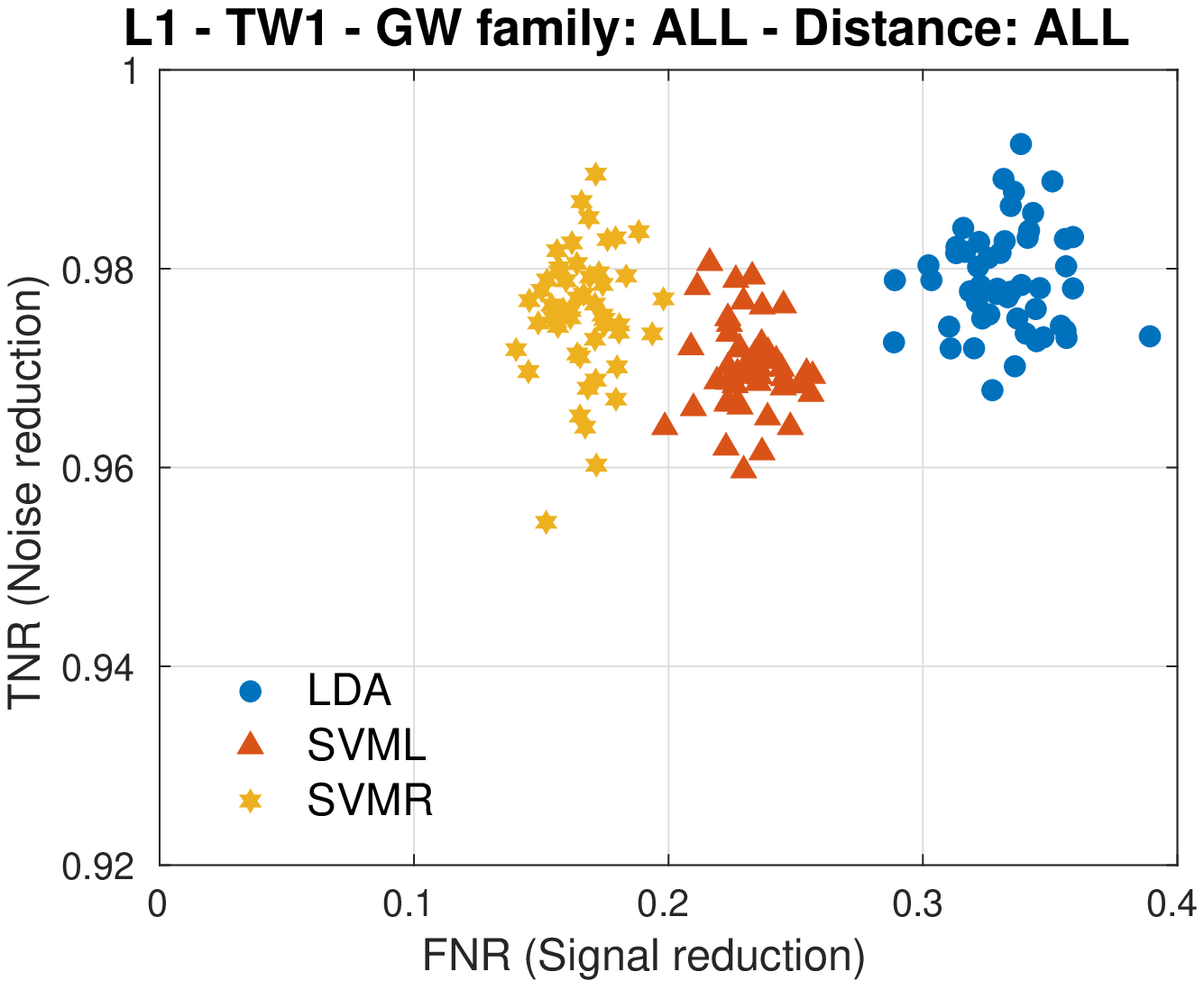}
    \\
    \\
    \includegraphics[width=0.28 \textwidth]{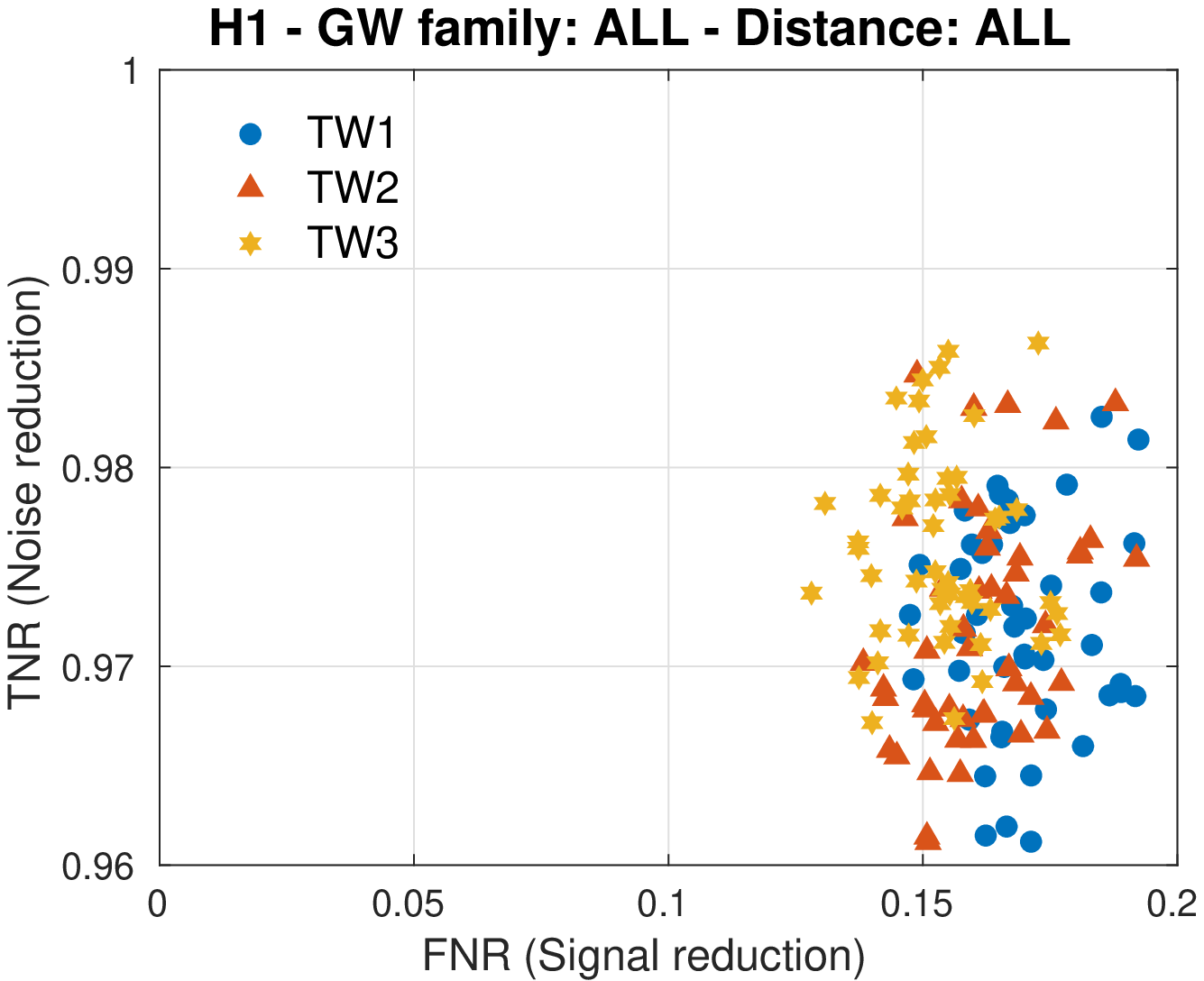}
    & 
    \includegraphics[width=0.28 \textwidth]{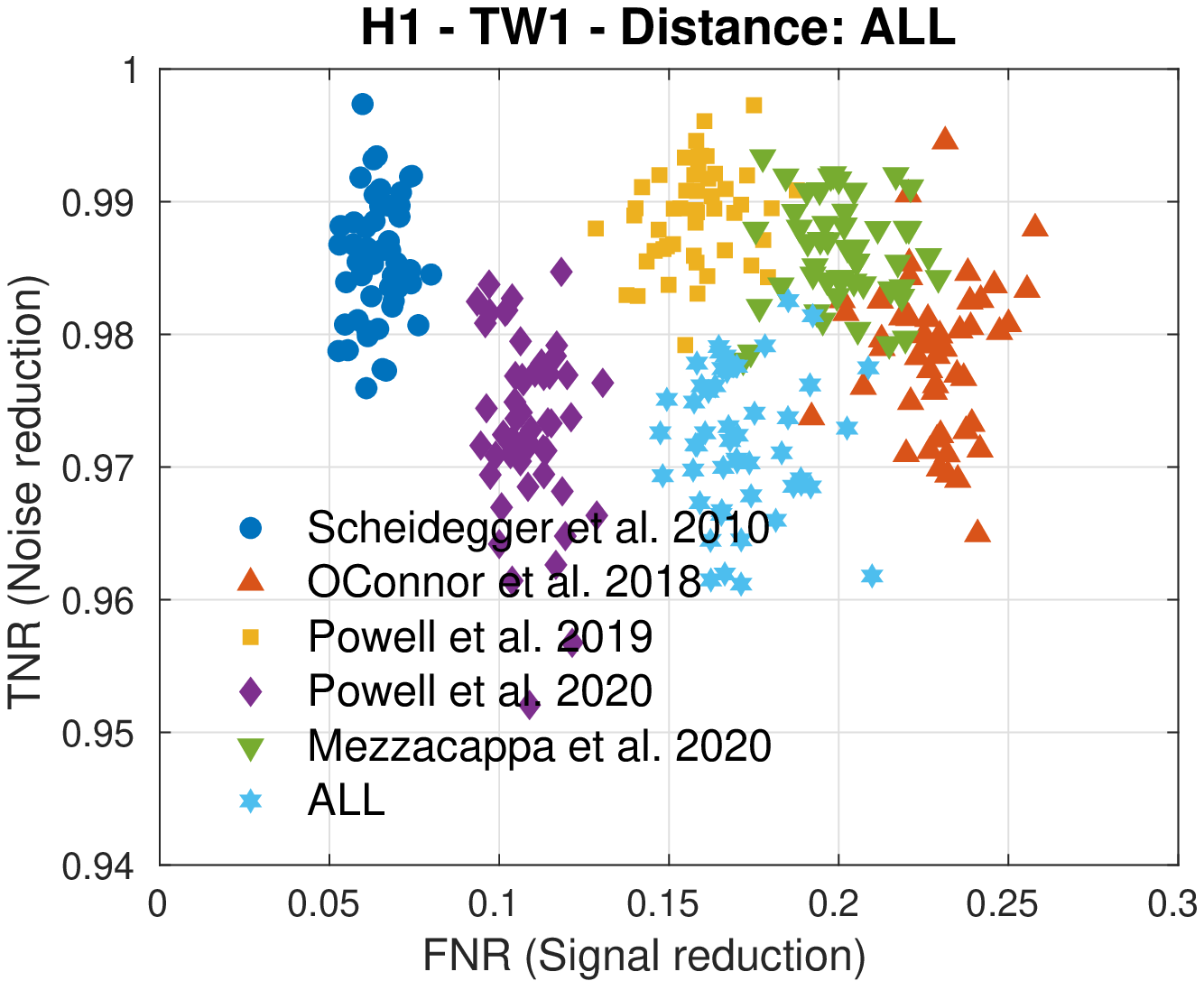}
    & 
    \includegraphics[width=0.305 \textwidth]{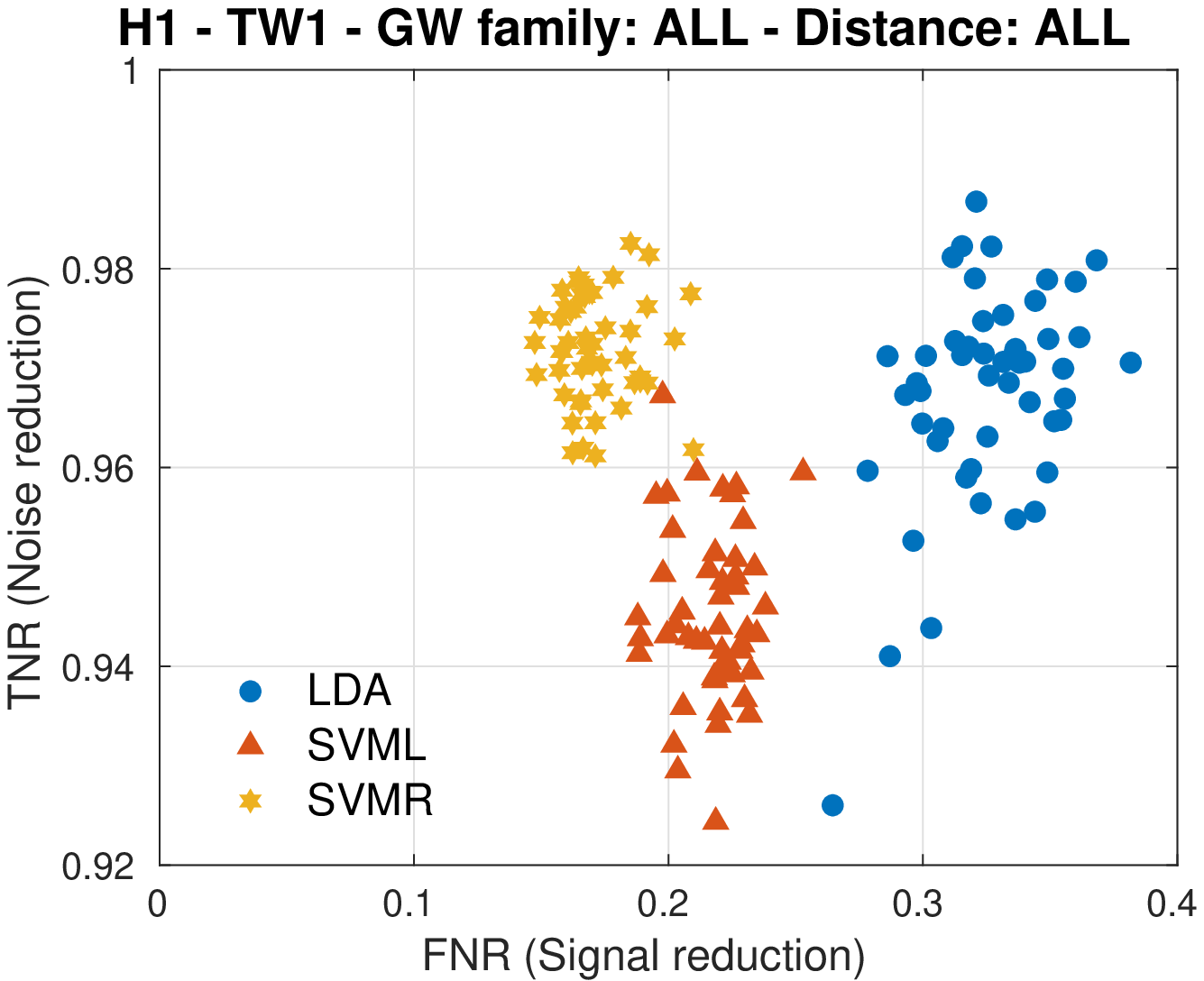}
    \\
    \\
    \includegraphics[width=0.28 \textwidth]{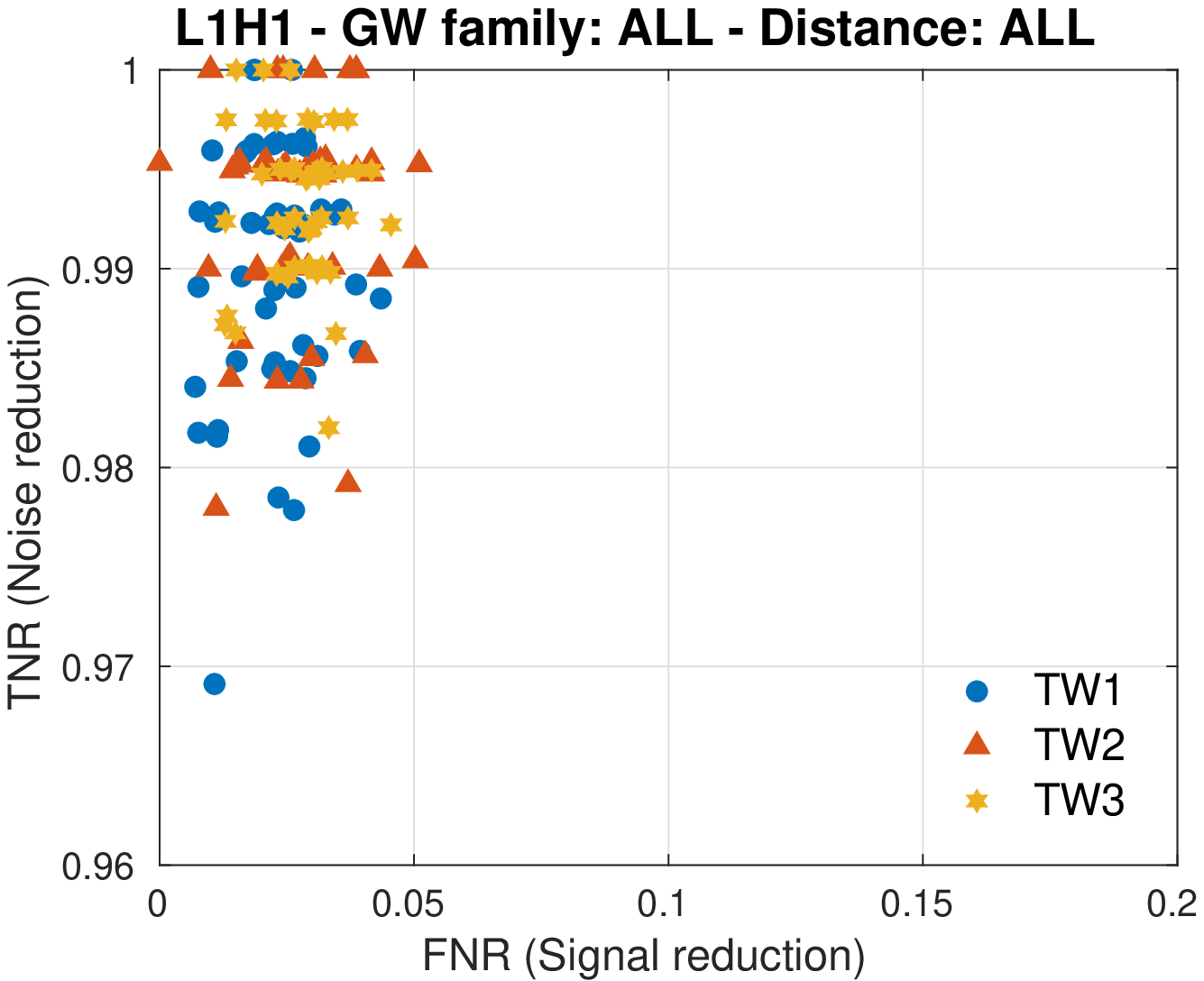}
    & 
    \includegraphics[width=0.28 \textwidth]{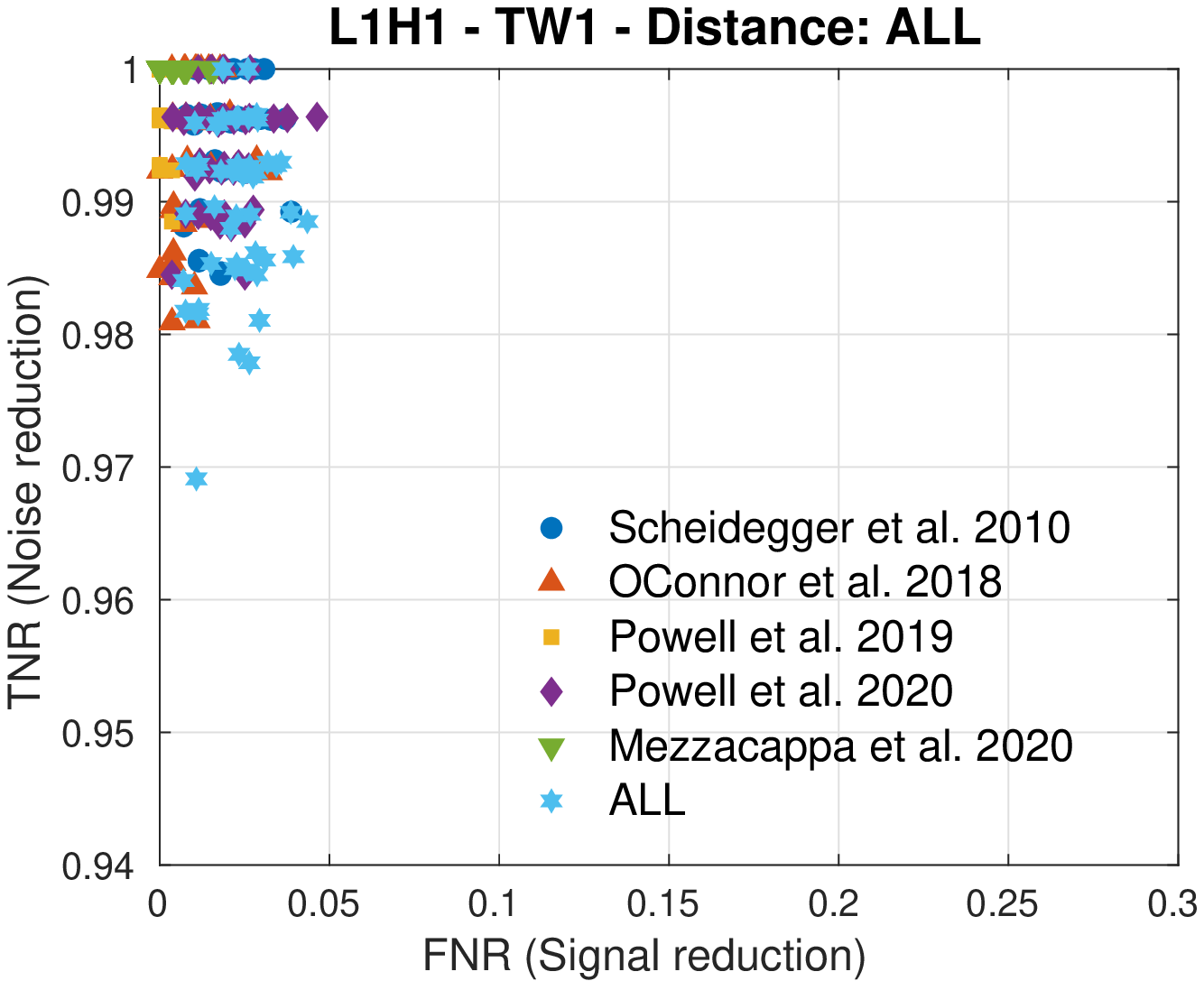}
    & 
    \includegraphics[width=0.31 \textwidth]{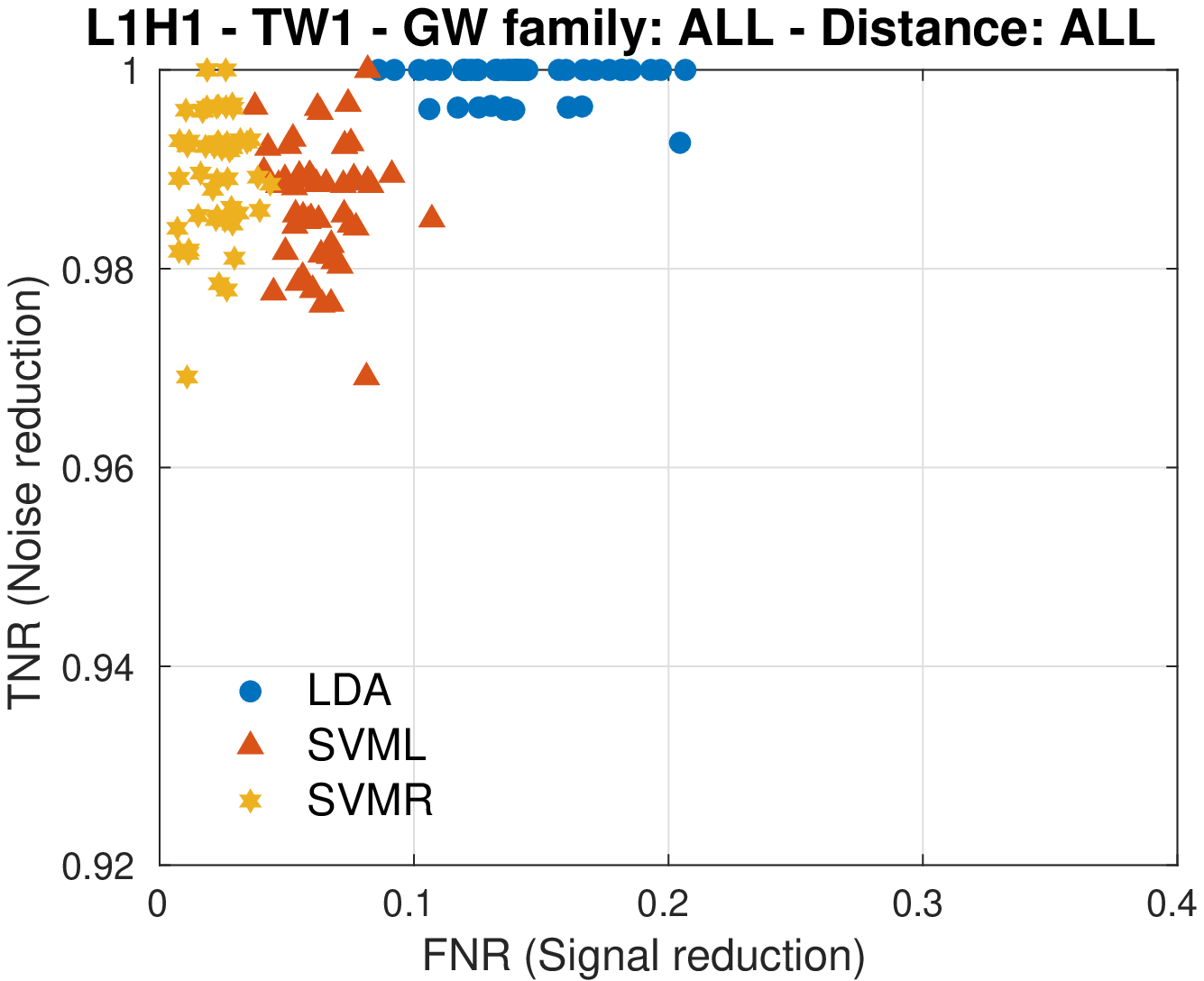}
    \\
    \\
    (a) & (b) & (c)
    \end{tabular}
    \caption{
    Distribution of classification metrics (TNR and FNR) obtained with (a) the three different CCSNe time windows, (b) each family of CCSNe GW, and (c) the three classification algorithms. These results are for the case of the single detector network L1 (upper panel), H1 (central panel) and L1H1 (lower panel).
    }
    \label{fig:CV_Compare}
\end{figure*}

\subsection{Training/Testing with data from the same time window}
As a first step to investigate the classification between noise and signal triggers we perform both the training and testing with information from the same stretch of strain data, i.e., noise and signal triggers used to train and test the classification model are from the same background and simulation analyses.
To do so, a cross-validation process was employed to assess classification performance.
The process was implemented as follows:
$(i)$ randomly split the dataset of noise and signal triggers into $K$ non-overlapping subsets or folds;
$(ii)$ use the data from $K-1$ subsets for training and the data from the remaining subset for testing (note that data for training and testing are always mutually exclusive);
$(iii)$ use the training set to calculate the parameters of the classification model;
$(iv)$ fed the classification model with all the triggers from the test set and compute performance metrics by comparing the output labels provided by the classifier with the corresponding true labels;
$(v)$ repeat steps $(ii)$ to $(iv)$ until all the $K$ combinations of train and test data are exhausted.

This cross-validation process with $K=5$ was repeated 10 times to be able to compute distributions of the classification performance metrics.
Prior to the training of the classification model, the number of noise and signal triggers in the training set is balanced to avoid overfitting to one of the classes, moreover, the training data was normalized according to $x_i = (x_i-\mu_i)/\sigma_i$, $(i=1,2,\cdots,N_f)$ where $\mu_i$ and $\sigma_i$ are the mean and standard deviation of the $i$-th feature which are computed exclusively from training data. The normalization is later applied to to each feature vector in the test set.

To assess performance the following metrics were computed: true negative rate (TNR) or specificity (i.e., correct classification percentage of noise events) and false negative rate (FNR) (i.e., percentage of signal triggers incorrectly classified as noise triggers).
TNR is expected to be large to reduce the FAR, while FNR is expected to be small to not affect the efficiency.
%
%
%
Classification analyses were carried out separately for each variant of:
\begin{itemize}
    \item \textit{Detector network:} L1, H1 and L1H1.
    \item \textit{Time window of strain data:} TW1, TW2, and TW3 (see table~\ref{Table:CCSNe_TimeWindows}). 
    \item \textit{Distance:} 1.00, 1.33, 1.78, 2.37, 3.16, 4.22, 5.62, 7.5, 10kpc, and all those distances together.
    \item \textit{Family of CCSNe GW:} \textit{Scheidegger et al. 2010}, \textit{O'connor et al. 2018}, \textit{Powell et al. 2019}, \textit{Powell et al. 2020}, and \textit{Mezzacappa et al. 2020}) and from all those families together.
    \item \textit{Classification models:} LDA, SVML, SVMR.
\end{itemize}

\begin{table*}
\caption{
Percentage of noise and signal reduction obtained by the three detector networks for each familiy of CCSNe GW waveforms and for several distances of signal triggers. These results are dataset of noise and signal triggers extracted from the time window TW1. For the two detector network there are missing results because it was not possible to train and test the classification model since no signal triggers were obtained in the simulation analysis.}
\begin{tabular*}{\textwidth} {@{\extracolsep{\textwidth minus \textwidth}} c|cccccccccccc} 
\hline
\hline
\multicolumn{1}{c|}{ }
& \multicolumn{2}{c}{1kpc}
& \multicolumn{2}{c}{2.37kpc}
& \multicolumn{2}{c}{4.22kpc}
& \multicolumn{2}{c}{7.5kpc}
& \multicolumn{2}{c}{10kpc}
& \multicolumn{2}{c}{All} 
\\
 &Noise&Signal & Noise&Signal &Noise&Signal & Noise&Signal & Noise&Signal & Noise&Signal  \\
\hline
\hline
\multicolumn{1}{c}{  } & \multicolumn{12}{c}{ } \\
\multicolumn{1}{c}{L1} & \multicolumn{12}{c}{ } \\
\hline
Scheidegger at al. 2010 & $99.6$ & $5.4$ & $99.6$ & $9.5$ & $99.4$ & $13.4$ & $99.0$ & $16.5$ & $98.7$ & $18.5$ & $99.0$ & $7.0$ \\
O'connor at al. 2018    & $97.0$ & $20.9$ & $93.4$ & $35.7$ & $91.9$ & $39.0$ & $90.9$ & $39.3$ & $91.5$ & $39.0$ & $97.5$ & $24.3$ \\
Powell at al. 2019      & $99.4$ & $16.2$ & $98.4$ & $18.5$ & $96.7$ & $22.1$ & $93.7$ & $29.6$ & $92.5$ & $34.5$ & $99.5$ & $15.8$ \\
Powell at al. 2020      & $98.8$ & $8.2$ & $97.5$ & $12.8$ & $96.7$ & $16.5$ & $95.8$ & $20.3$ & $95.5$ & $22.4$ & $97.9$ & $10.4$ \\
Mezzacappa at al. 2020  & $99.2$ & $18.1$ & $97.2$ & $23.7$ & $94.0$ & $35.8$ & $93.5$ & $39.5$ & $93.0$ & $39.1$ & $98.9$ & $20.8$ \\
All                     & $97.6$ & $16.1$ & $96.7$ & $23.0$ & $95.3$ & $25.7$ & $93.4$ & $29.7$ & $93.8$ & $32.2$ & $97.5$ & $16.6$ \\
\hline
\multicolumn{1}{c}{  } & \multicolumn{12}{c}{ } \\
\multicolumn{1}{c}{H1} & \multicolumn{12}{c}{ } \\
\hline
Scheidegger at al. 2010 & $99.0$ & $4.3$ & $98.9$ & $9.1$ & $98.8$ & $12.6$ & $98.5$ & $15.2$ & $98.2$ & $16.5$ & $98.6$ & $6.4$ \\
O'connor at al. 2018    & $96.6$ & $23.3$ & $93.4$ & $32.7$ & $93.8$ & $33.2$ & $93.5$ & $33.0$ & $93.8$ & $33.7$ & $97.8$ & $22.9$ \\
Powell at al. 2019      & $99.4$ & $15.1$ & $97.8$ & $18.7$ & $96.2$ & $24.5$ & $93.5$ & $31.4$ & $92.3$ & $33.0$ & $98.9$ & $15.9$ \\
Powell at al. 2020      & $97.9$ & $7.7$ & $96.3$ & $14.0$ & $94.4$ & $16.6$ & $93.7$ & $19.2$ & $93.7$ & $20.9$ & $97.3$ & $10.9$ \\
Mezzacappa at al. 2020  & $99.2$ & $18.2$ & $96.2$ & $27.3$ & $93.5$ & $33.2$ & $93.8$ & $33.3$ & $93.1$ & $33.5$ & $98.6$ & $20.1$ \\
All                     & $97.1$ & $17.4$ & $95.6$ & $22.1$ & $94.1$ & $25.0$ & $93.2$ & $27.4$ & $93.1$ & $28.2$ & $97.2$ & $17.2$ \\
\hline
\multicolumn{1}{c}{  } & \multicolumn{12}{c}{ } \\
\multicolumn{1}{c}{L1H1} & \multicolumn{12}{c}{ } \\
\hline
Scheidegger at al. 2010 & $98.6$ & $2.2$ & $99.7$ & $1.0$ & $99.9$ & $0.9$ & $99.7$ & $0.1$ & $99.7$ & $0.2$ & $99.5$ & $1.9$ \\
O'connor at al. 2018    & $98.7$ & $1.3$ & $-$ & $-$ & $-$ & $-$ & $-$ & $-$ & $-$ & $-$ & $99.4$ & $1.1$ \\
Powell at al. 2019      & $100.0$ & $0.3$ & $99.3$ & $0.3$ & $99.0$ & $0.3$ & $95.9$ & $0.8$ & $-$ & $-$ & $99.8$ & $0.1$ \\
Powell at al. 2020      & $99.8$ & $0.1$ & $99.3$ & $1.5$ & $98.5$ & $3.2$ & $98.7$ & $5.0$ & $98.2$ & $7.0$ & $99.4$ & $1.9$ \\
Mezzacappa at al. 2020  & $100.0$ & $0.4$ & $99.3$ & $0.5$ & $-$ & $-$ & $-$ & $-$ & $-$ & $-$ & $100.0$ & $0.4$ \\
All                     & $99.5$ & $0.9$ & $99.0$ & $1.3$ & $98.7$ & $3.3$ & $98.2$ & $7.5$ & $96.1$ & $8.5$ & $99.0$ & $2.2$ \\
\hline
\hline
\end{tabular*}
\label{Table:CV_NoiseAndSignalReduction}
\end{table*}

Figure~\ref{fig:CV_TNRandFNR}a shows the correct noise classification rate (TNR) and the incorrect signal classification rate (FNR) obtained with the dataset from the time window TW1 separately with detector networks H1, L1, and L1H1.
These results are for the case of signal triggers from all distances and from all families of CCSNe GW, which represents the most difficult situation for a classification model because they combine signal triggers with different SNR and with different GW signatures.
The percentage of noise triggers that are correctly classified is on average $97.6\pm0.6$, $97.2\pm0.5$ and $99.4\pm0.6$\% for detector networks L1, H1 and L1H1, respectively.
This indicates very high specificity (nearly 100\%) with very low variability (less than 0.6\%) irrespective of the number of detectors in the network.
On the contrary, signal triggers are lost, on average, $16.6\pm1.2$, $17.2\pm1.4$ and $2.2\pm0.9$\% for detector networks L1, H1 and L1H1, respectively.
These results shows a very high noise reduction irrespective of the detector network and low signal lost especially for the detector network with two interferometers.

To examine the effect of the distance, figure~\ref{fig:CV_TNRandFNR}b shows the classification results obtained with signal triggers from each distance separately.
As the distance increases, correct noise classification (TNR) decreases while signal misclassification (FNR) increases and the proportion of lost signals increases faster than the decrease in correct noise classification. 
This is due to the fact that as the distance increases the SNR of the GW signals decreases, and therefore it becomes more difficult to discriminate between noise and signal triggers because they tend to have similar characteristics.

The results presented in figure~\ref{fig:CV_TNRandFNR} a and b include signal triggers from all GW families considered herein, which represent the case with the more variability in the characteristics of signal triggers.
Similar results and observations were drawn with analysis carried out with datasets of noise and signal triggers obtained from time windows TW2 and TW3.
Altogether, these results shows that a better classification between noise and signal triggers with the network of two detectors L1H1, and the lower performance with the network of one detector H1.

Figure~\ref{fig:CV_Compare} shows the classification results from the cross-validation analysis for the three time windows, each family of CCSNe GW, and for the three classification algorithms.
First, with regard to the three stretches of strain data (Figure~\ref{fig:CV_Compare}a), high noise detection and low signal misclassification was obtained in the three cases, and there is a similar distribution of the performance metrics.
Second, there is some variable classification performance considering the family of CCSNe GW (Figure~\ref{fig:CV_Compare}b). In particular, using signal triggers from \textit{O'Connor et al. 2018} yielded to a lower noise detection and higher signal lost, while triggers from \textit{Scheidegger et al. 2010}, \textit{Powell et al. 2019} and \textit{Powell et al. 2020} resulted in the best performance with the higher TNR and lower FNR. Importantly, note that for the case of using signal triggers from all families combined the performance metrics are distributed within all families.
Finally, regarding the comparison between classification algorithms, all classifiers achieved similar and high rates of noise reduction, however, the rate of signal lost varied across them, in particular, SVMR presented the lower signal lost. In consequence, the non-linear SVMR classifier was selected to be used is the subsequent analyses.
The same behavior of the noise reduction and signal lost is observed in all detector networks (L1, H1 and L1H1), but the network of two detectors always presents the best performance

A summary of the cross-validation results with the percentage of noise and signal reduction for each network of detectors (L1, H1 and L1H1), for each family of CCSNe GW, for different distances, and for the first time window of data is presented in table~\ref{Table:CV_NoiseAndSignalReduction}.
These results shows that classification performance degrades as the distance increases, shows no large differences across families of CCSNe GW and is higher for the two detector network.

\begin{figure*}[t]
    \centering
    \begin{tabular}{ccc}
    \includegraphics[width=0.27 \textwidth]{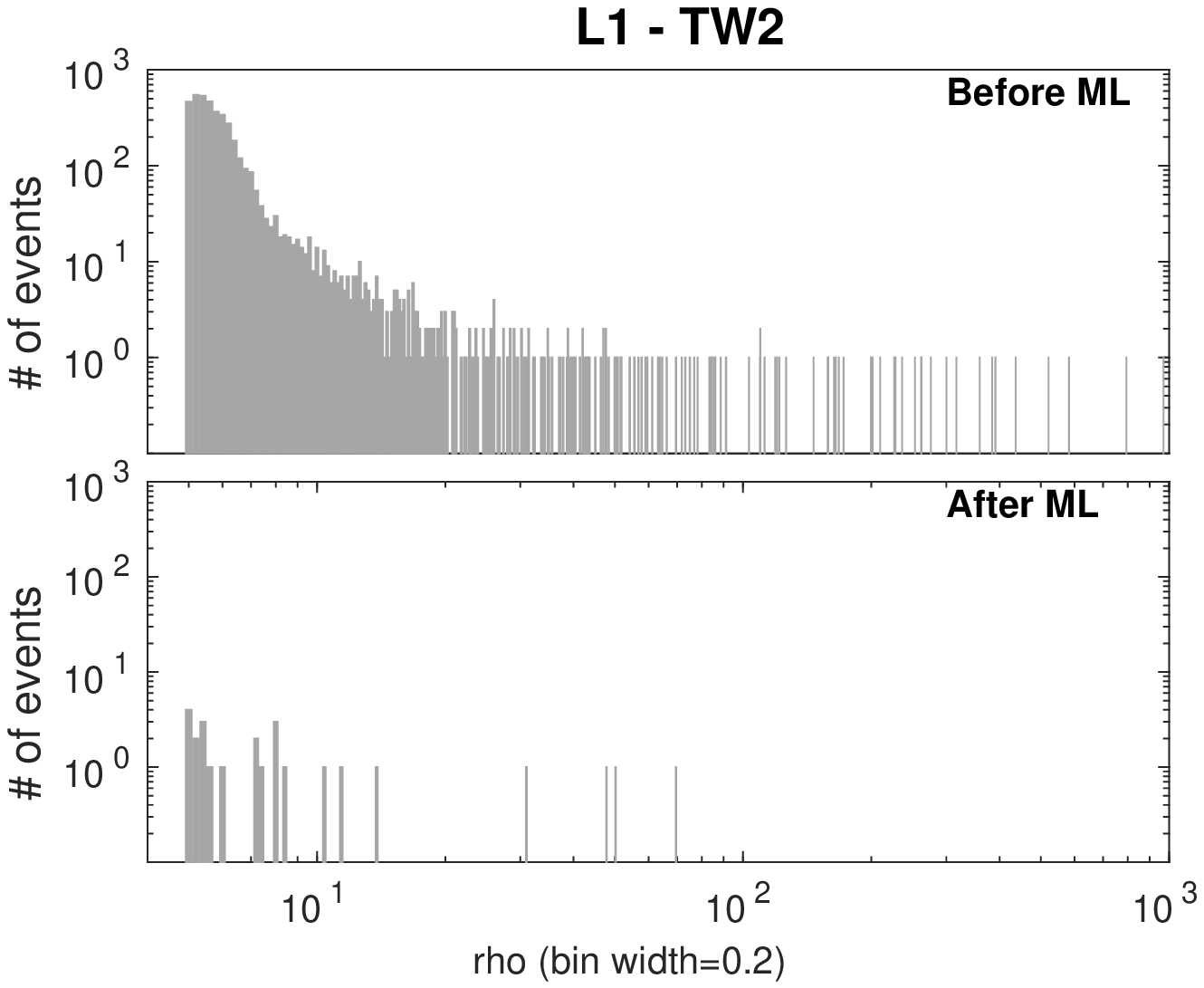}
    &
    \includegraphics[width=0.27 \textwidth]{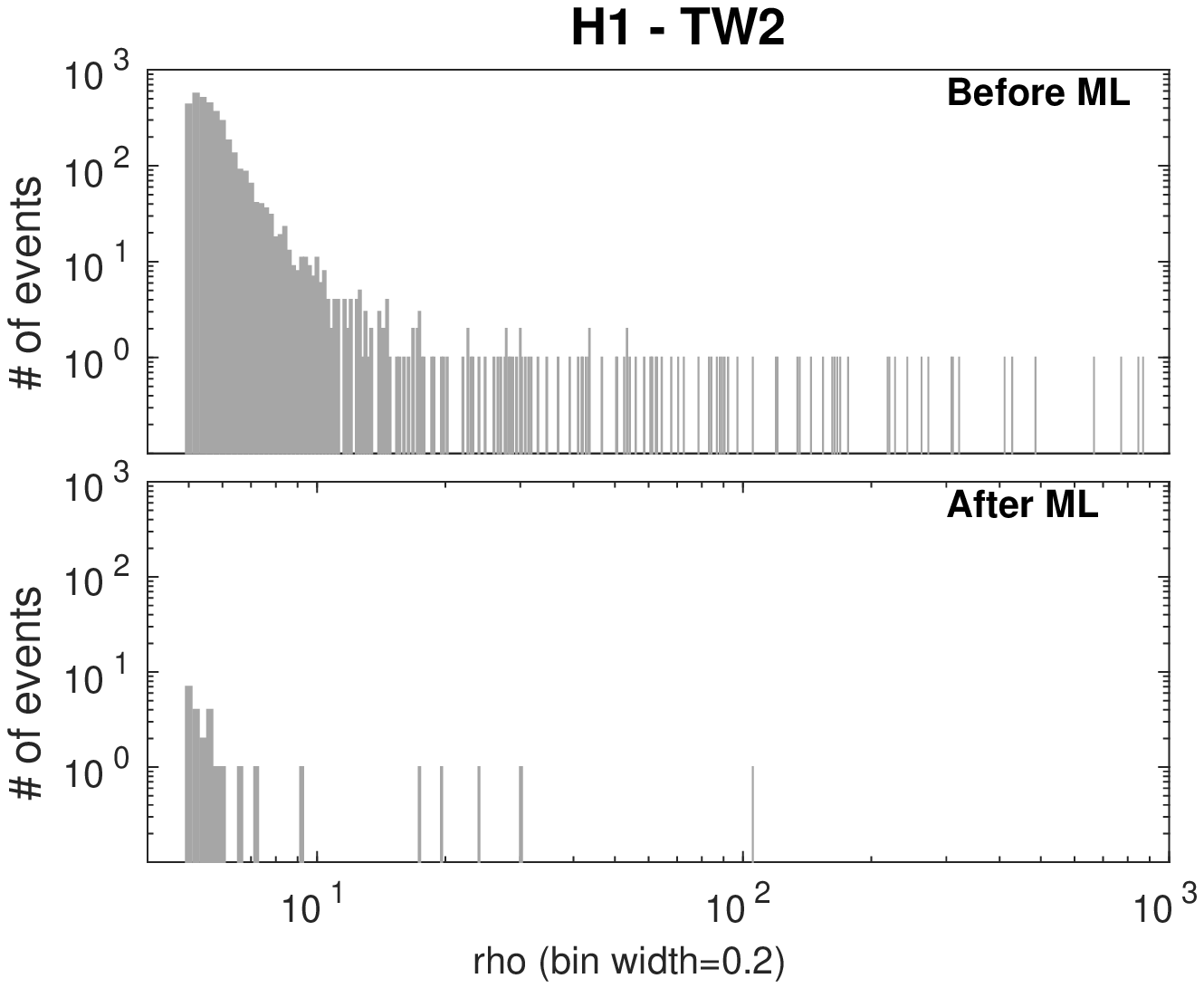}
    &
    \includegraphics[width=0.27 \textwidth]{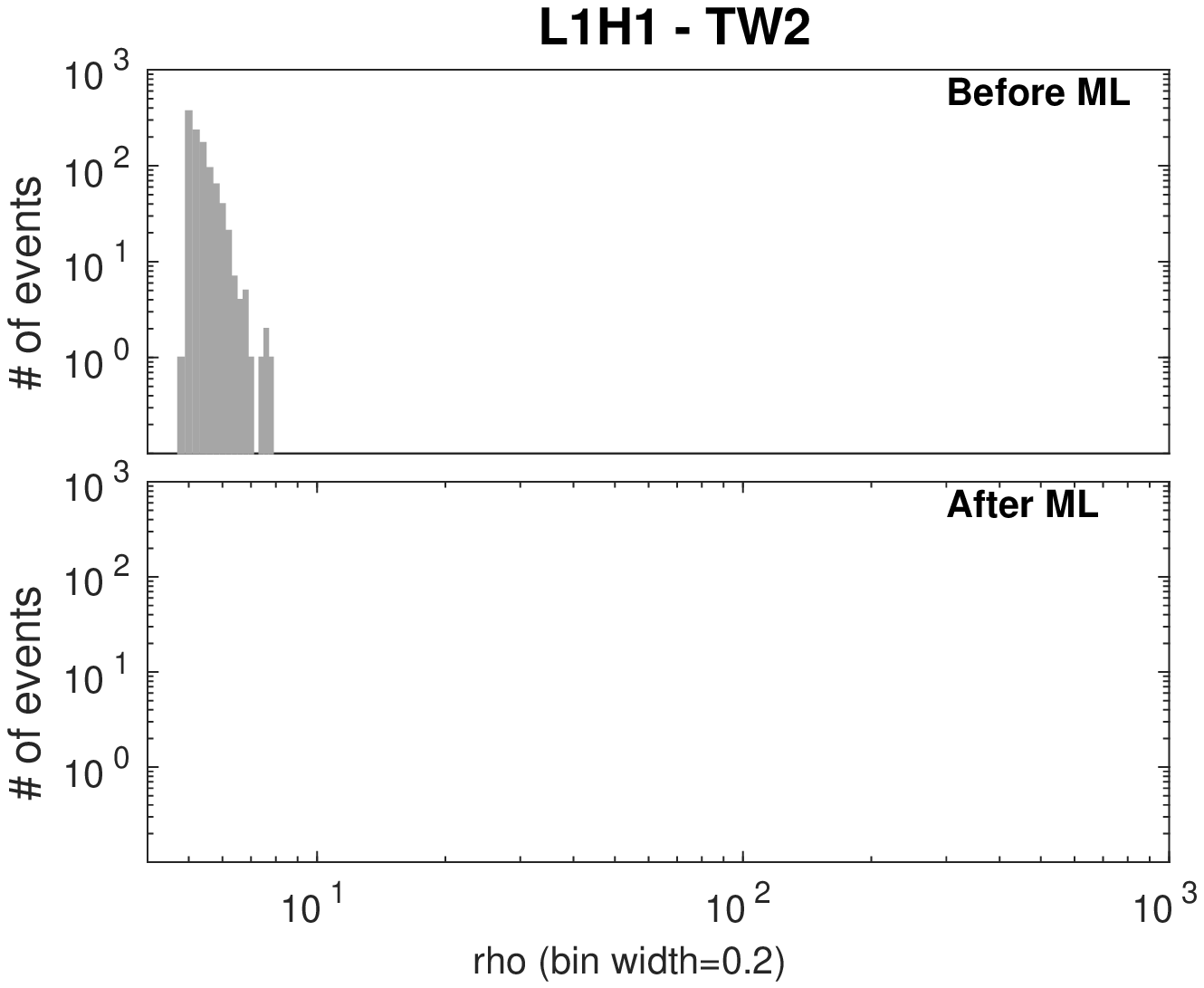}
    \\
    \\
    \\
    \includegraphics[width=0.3 \textwidth]{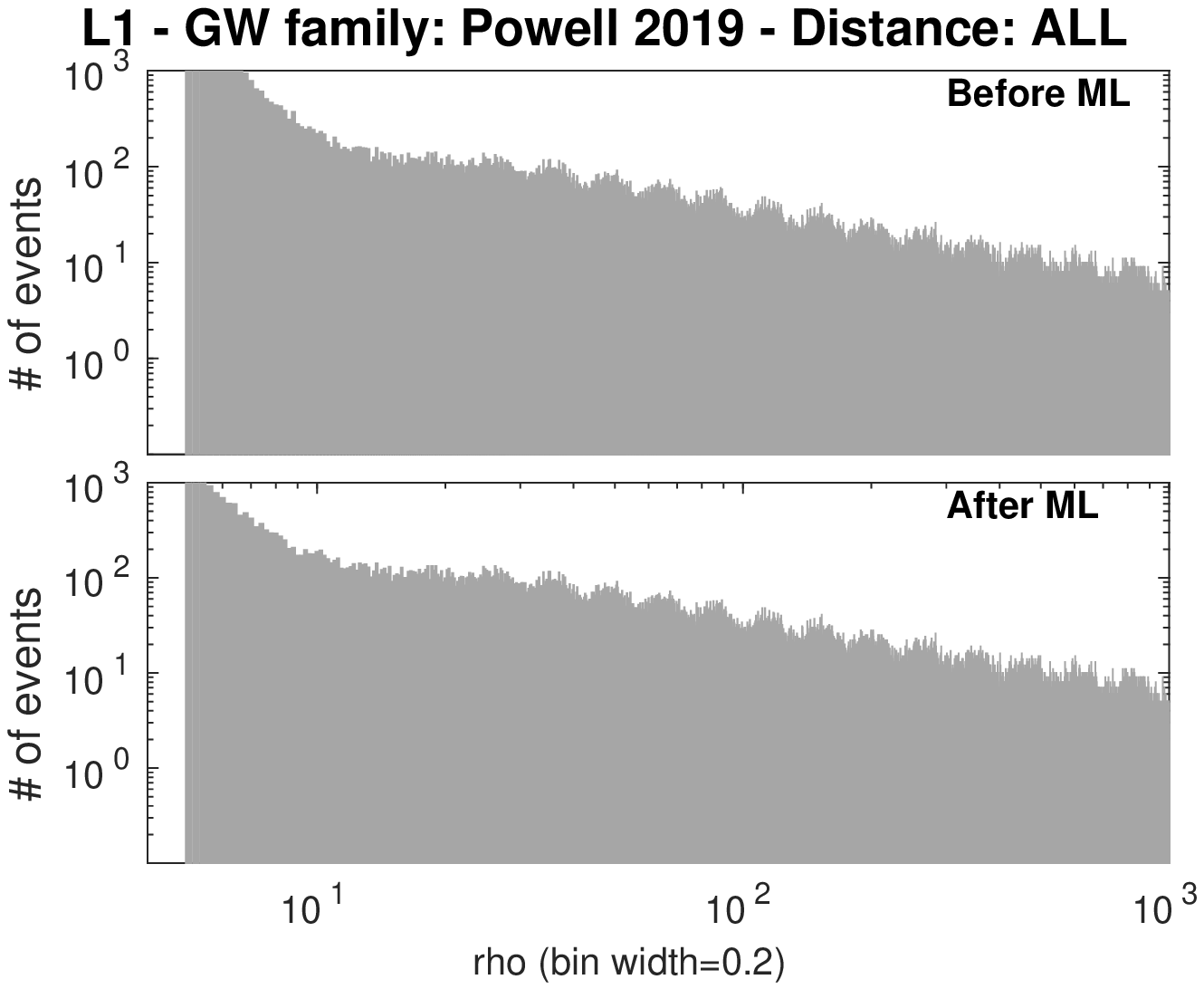}
    &
    \includegraphics[width=0.3 \textwidth]{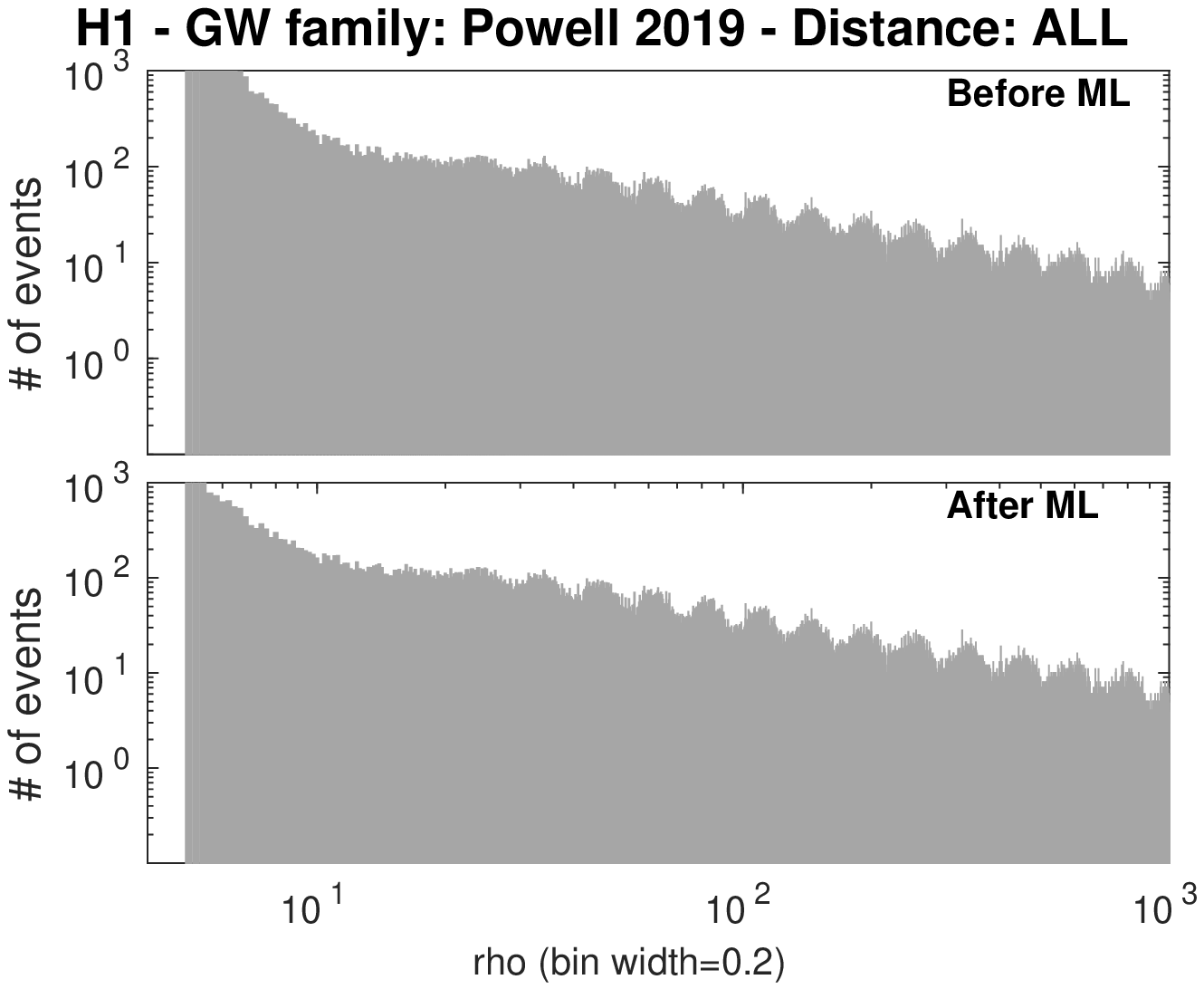}
    &
    \includegraphics[width=0.3 \textwidth]{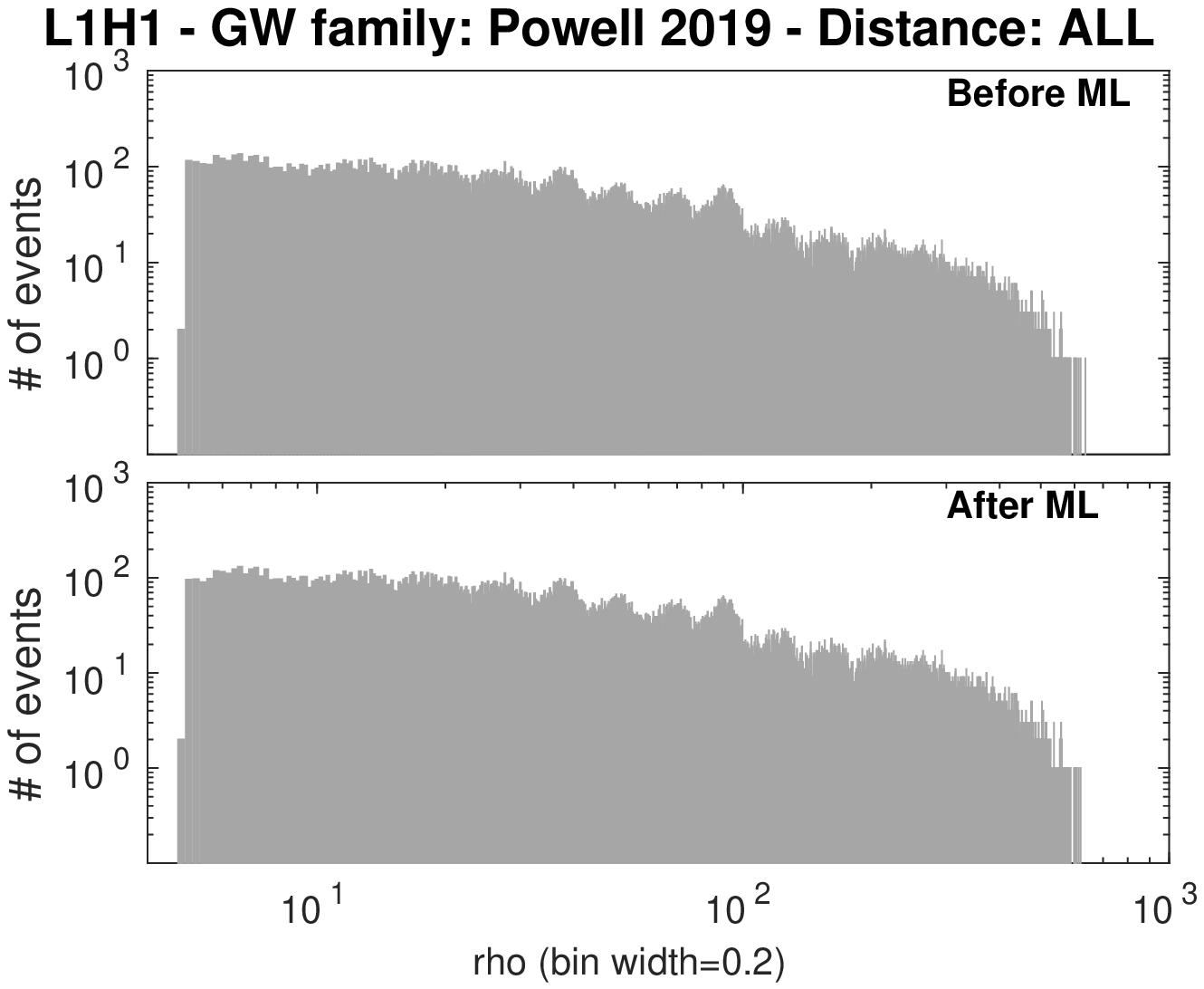}
    \\ 
    (a) & (b) & (c)
    \end{tabular}
    \caption{
    Distribution of noise and signal triggers before and after the application of the classification model with CCSNe GWs from \textit{Powell et al. 2019}.
    These results are for the cWB analysis in the time window TW2 for each detector network (a) L1, (b) H1, and (c) L1H1.
    For each detector, the classification model was tuned with the corresponding dataset of noise and signal triggers extracted from time window TW1 with signal triggers from all distances.
    Top panel shows the noise triggers while bottom panel shows the signal triggers.
    }
    \label{fig:OS_ResultsEventsBeforeAfter}
\end{figure*}

\subsection{Training/Testing with data from different time windows}
The single interferometer scenario highlight the problem of only having a small amount of live time to tune the parameters of the ML model.
This second study aimed to assess the actual improvement (i.e., noise reduction) and the potential drawback (i.e., signal reduction) given by the incorporation of the ML model as a follow-up method in cWB offline searches of GWs from CCSNe.
Here, the classification model is learned with a dataset obtained from a given cWB analysis, and then, it is used in a different cWB analysis.
This allows to effectively quantify the improvement in FAR and FAP, while also measuring the impact in the DE.
In addition, to assess the robustness to unknown GWs, the ML model is tuned using all but one of families of GW signals, and then, it is tested with the remaining family.

Once the classification model is applied to a offline cWB analysis, the following performance metrics are computed:
$(i)$ noise reduction rate (TNR), and signal reduction rate (FNR);
$(ii)$ FAR and FAP before and after the application for the ML model;
$(iii)$ DE before and after the application for the ML model.
Note that FAR, FAP and DE ``before" refer to the original search results provided by cWB, while FAR, FAP and DE ``after" refer to the results obtained with cWB in combination with the ML model.
As in the first study, analyses were carried out separately with each detector network (L1, H1, L1H1), signal triggers from each distance and from all distances combined.
%
%
%

The results presented below are for the specific case of training the ML model with triggers extracted from the cWB analysis in the time window TW1, whereas the model is applied to the cWB analysis carried out in the time window TW2 using exclusively CCSNe GWs from \textit{Powell et al. 2019}.
%
%
%
Firstly, figure~\ref{fig:OS_ResultsEventsBeforeAfter} shows the distribution of noise and signal events before and after the application of the ML model.
Noise events are significantly reduced with only a small proportion remaining after classification, indeed, noise reduction is 99.4\%, 99.2\%, and 100.0\%, for L1, H1 and L1H1, respectively.
At the same time, most of the signal events remains after the application of the classifier since there is low signal reduction of 15.7\%, 14.3\%, and 0.3\%, for L1, H1 and L1H1, respectively.
This shows that irrespective of the detector network, a high proportion of the noise triggers are removed while a small proportion of signal triggers are lost.
In addition, the higher noise reduction and the lower signal lost were obtained with the two detector network.
%

\begin{figure*}[t]
    \centering
    \begin{tabular}{ccc}
    \includegraphics[width=0.26 \textwidth]{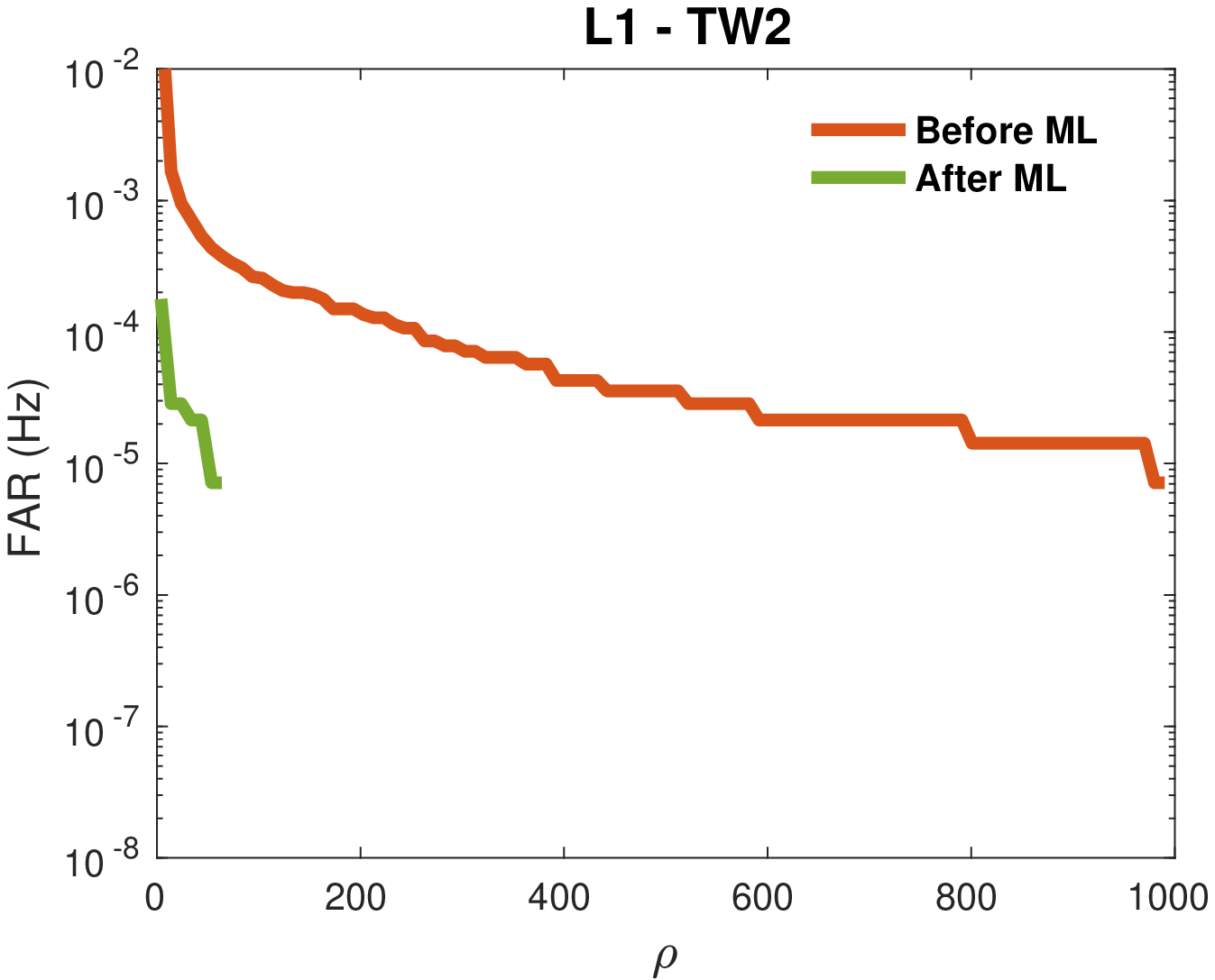}
    & 
    \includegraphics[width=0.26 \textwidth]{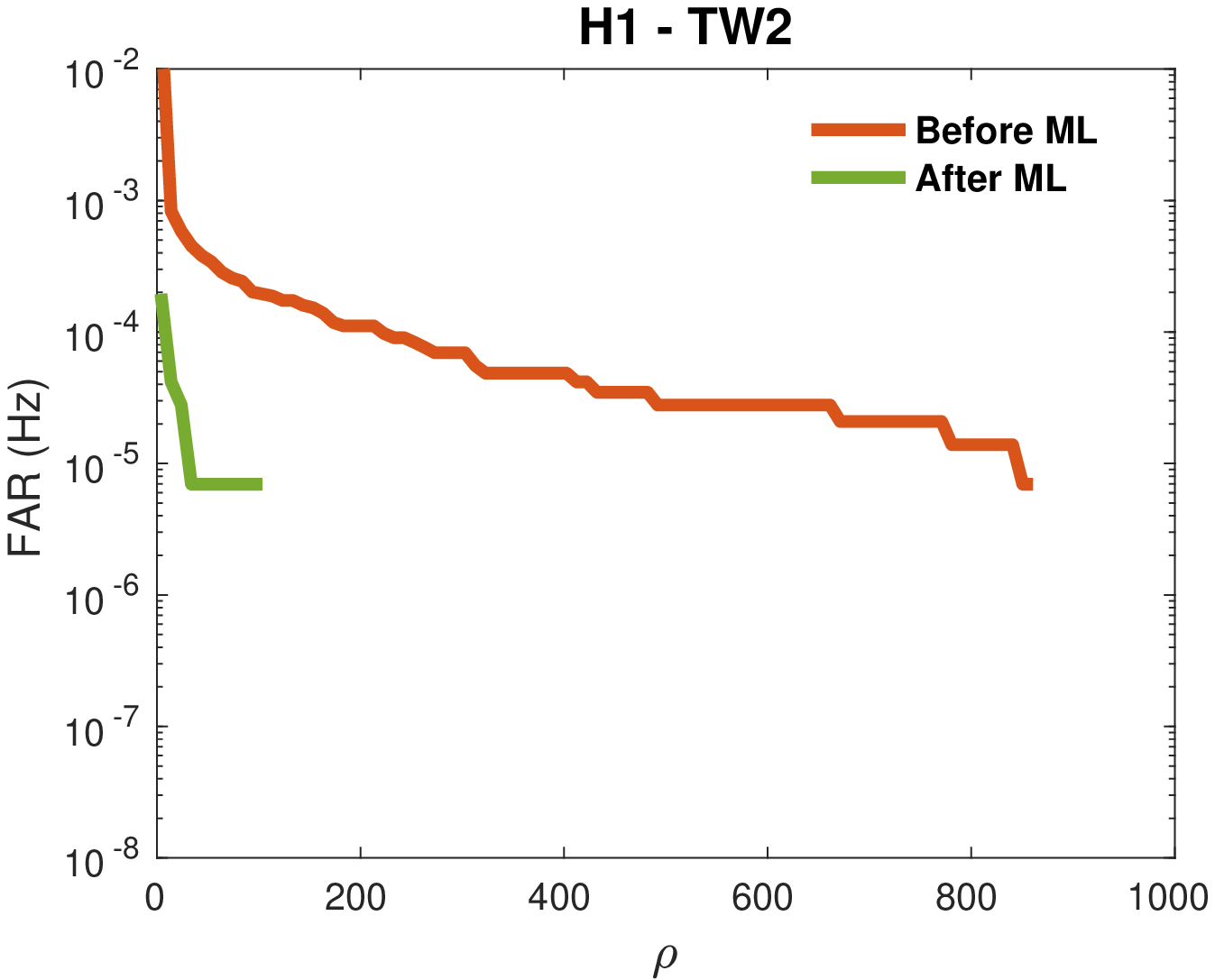}
    & 
    \includegraphics[width=0.25 \textwidth]{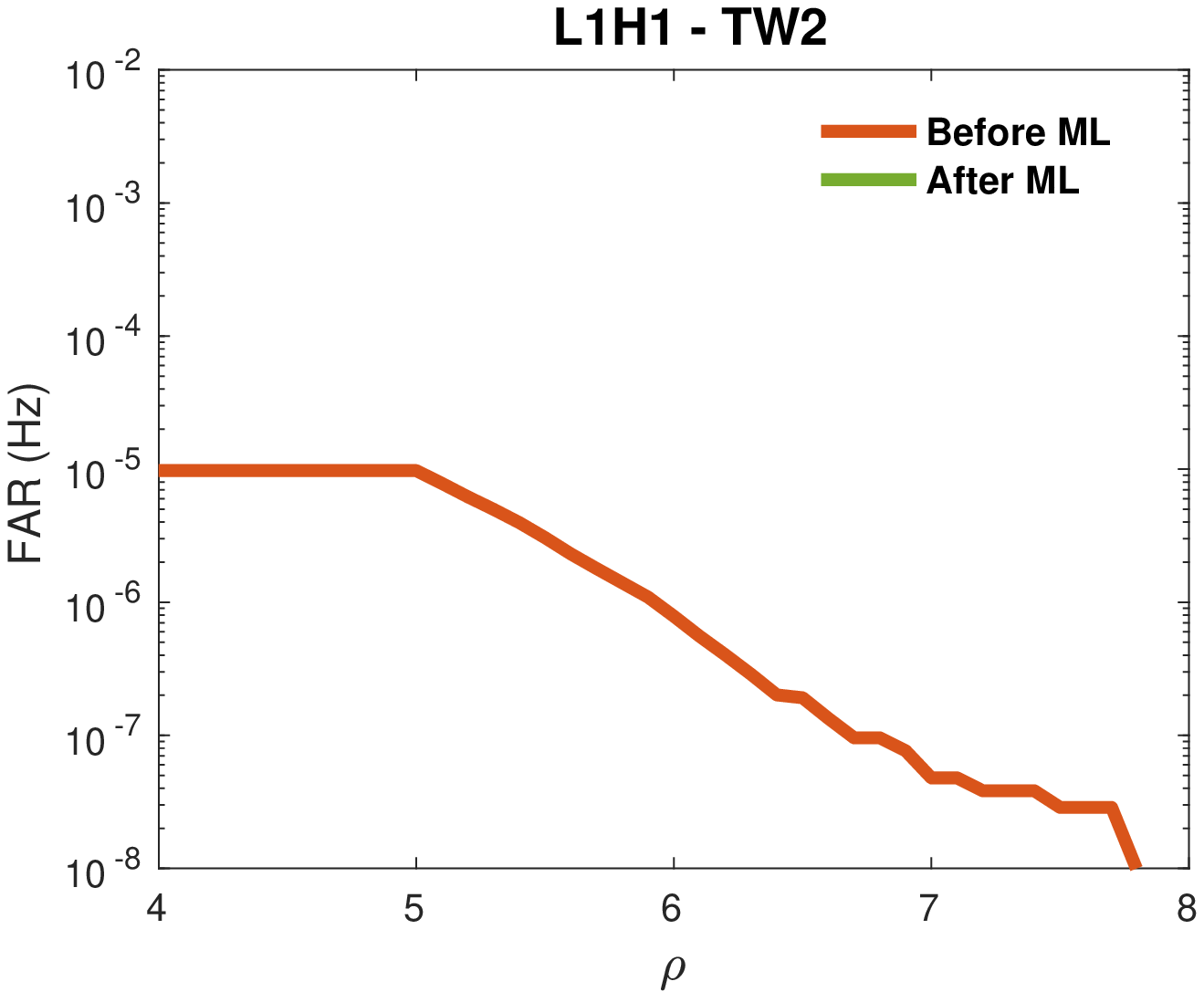}
    \\
    \\
    \includegraphics[width=0.26 \textwidth]{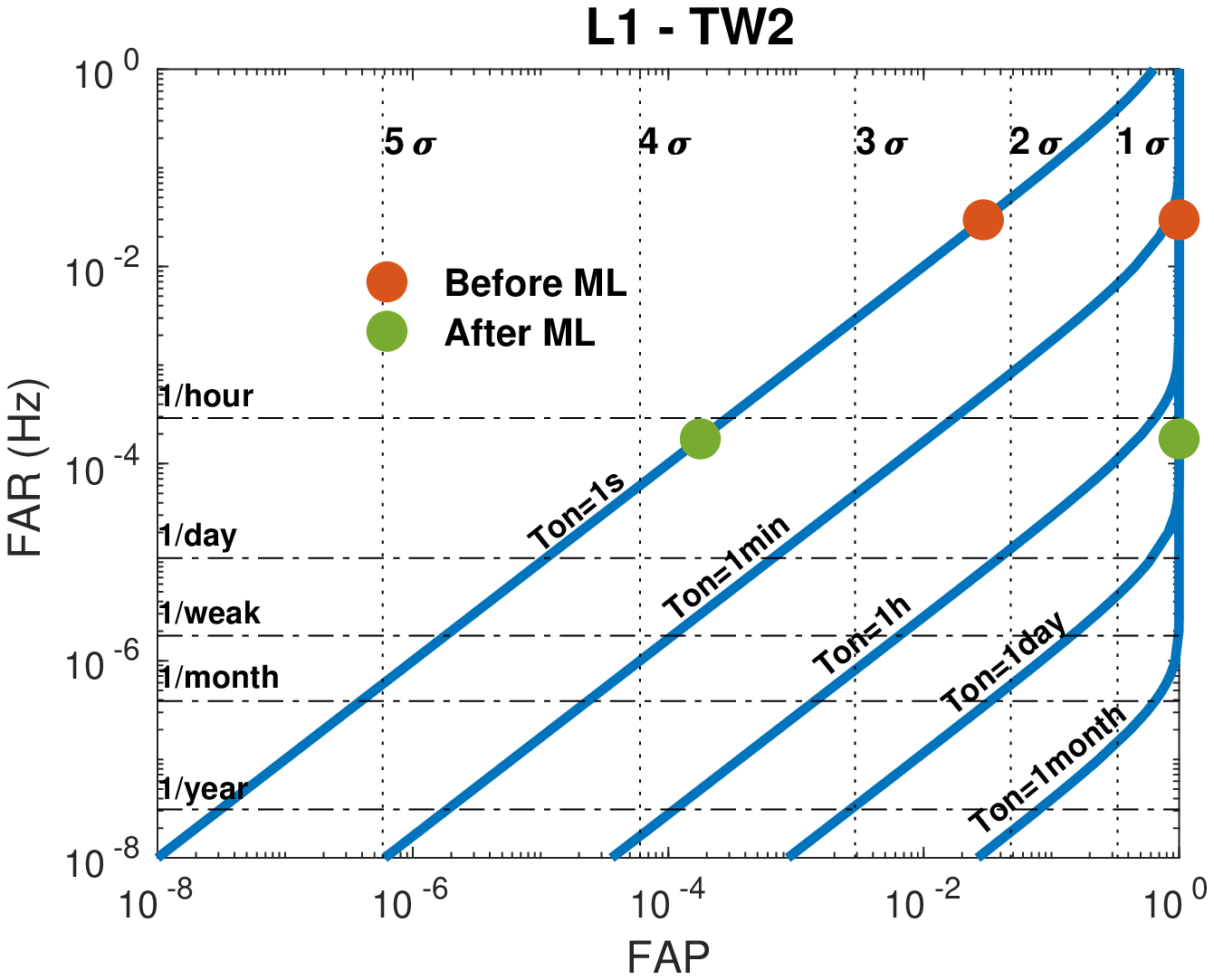}
    & 
    \includegraphics[width=0.26 \textwidth]{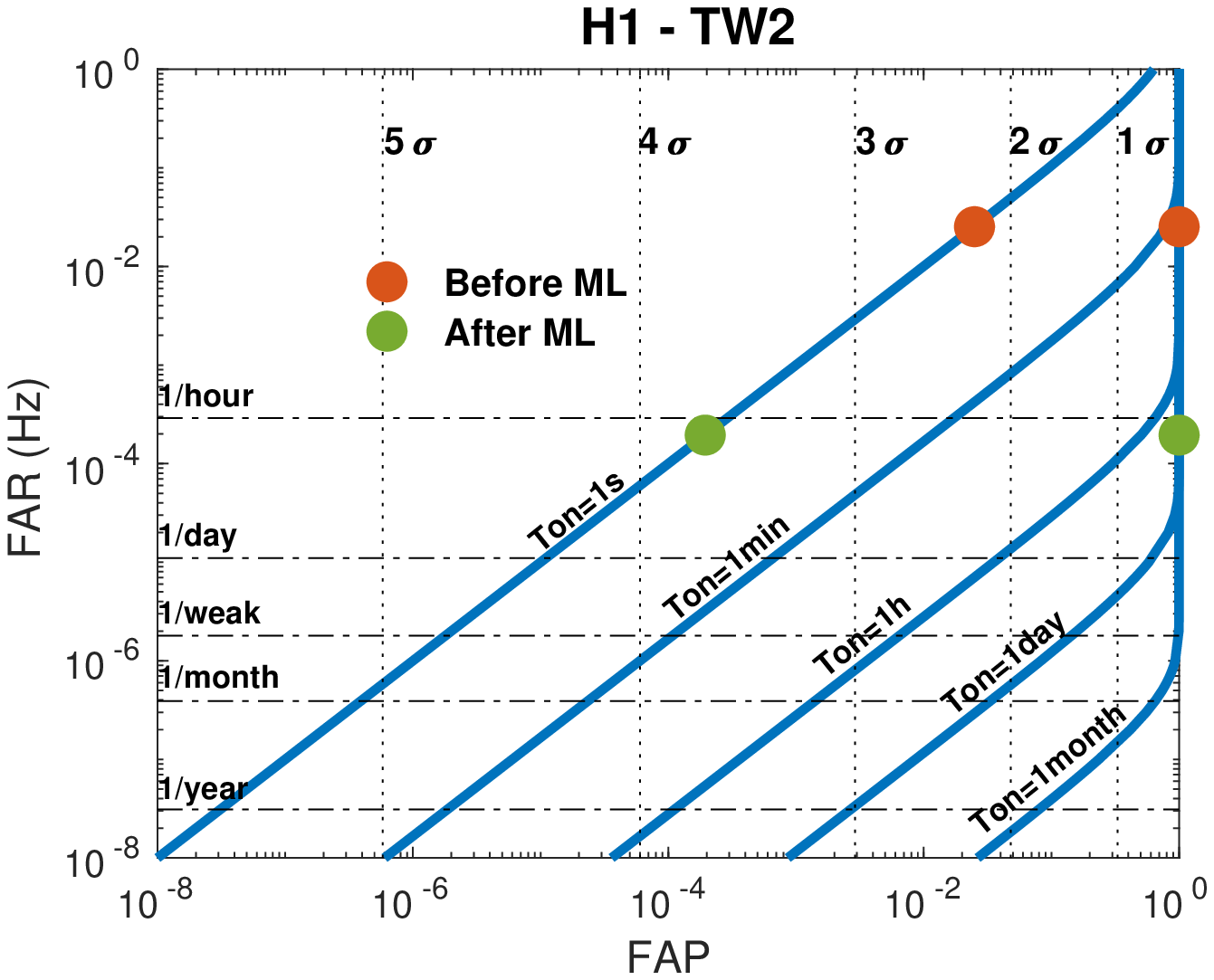}
    & 
    \includegraphics[width=0.26 \textwidth]{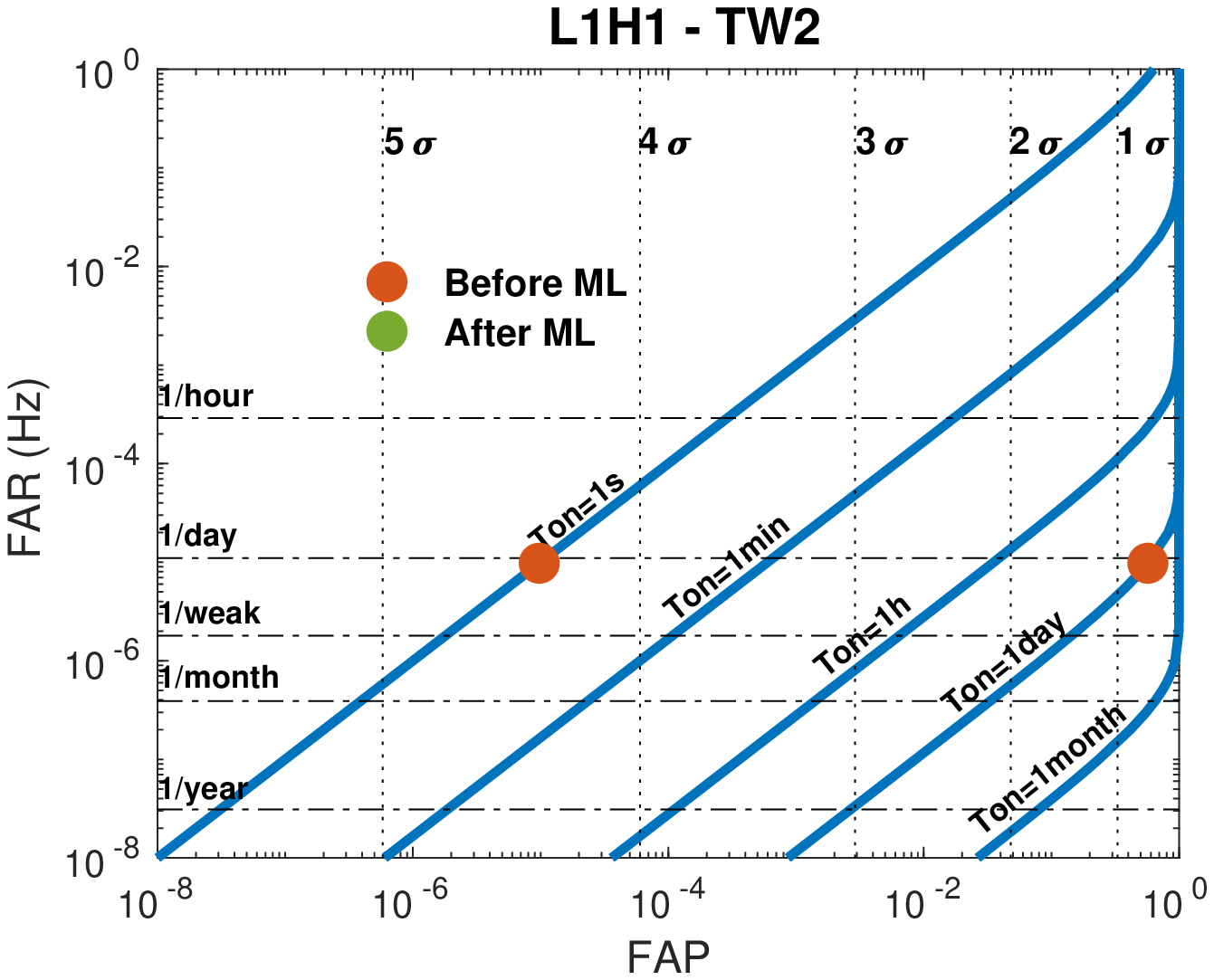}
    \\
    (a) & (b) & (c)
    \end{tabular}
    \caption{
    Results of the cWB background search in the time window TW2 before (red) and after (green) the application of the classification model.
    Top panel shows the false alarm rate (FAR) versus the signal-to-noise ratio ($\rho$), and bottom panel shows the FAP versus false alarm rate (FAR) for the two representative cases in the search of GWs from CCSNe (on-source of 1s resembling a neutrino-driven search, and a on-source of 1day representing a optical-targeted search).
    These results are for each detector network, (a) L1, (b) H1, and (c) L1H1.
    For each detector network, the classification model was trained with the corresponding dataset of triggers extracted from the time window of TW1 and with signal triggers from all distances.
    }
    \label{fig:OS_FARandFAP_BeforeAfter}
\end{figure*}
Secondly, figure~\ref{fig:OS_FARandFAP_BeforeAfter} shows the FAR versus $\rho$ (or SNR for the case of single interferometer), and the FAP versus FAP before and after the application of the classifier. These are crucial results to determine the actual improvement produced by the incorporation of the classifier as a follow-up method.
On the one hand, figure~\ref{fig:OS_FARandFAP_BeforeAfter}a shows a clear FAR reduction, especially in low values of $\rho$.
For the specific operating point of the cWB search (i.e., network $\rho=5$), the FAR before/after the classification model is $2.97\times10^{-2}$/$1.78\times10^{-4}$, $2.53\times10^{-2}$/$1.95\times10^{-4}$, and $9.76\times10^{-6}$/$9.58\times10^{-9}$ for L1, H1, and L1H1, respectively. 
This indicates that the FAR is effectively lowered in a factor of up to $\sim 170$, $\sim 130$, and $\sim1019$.
On the other hand, figure~\ref{fig:OS_FARandFAP_BeforeAfter}b shows the false alarm probability (FAP) before and after for on-source windows of 1s (representing a potential neutrino driven search) and of 1day (representing a optically targeted search).
For the single detector network, the FAP improvement is about $1.5\sigma$ in the case of a on-source window of 1s, while no improvement was achieved in the case of a on-source window of 1day.
In contrast, with the two detector network there is FAP improvement is about $1\sigma$.
We stress that these results represent the most difficult situation to tune and to test the classification model since it involves signal triggers of different distances, and because, the ML model is recognizing unknown GW waveforms that were not employed in the training.

%

\begin{figure*}[t]
    \centering
    \begin{tabular}{ccc}
    \includegraphics[width=0.26 \textwidth]{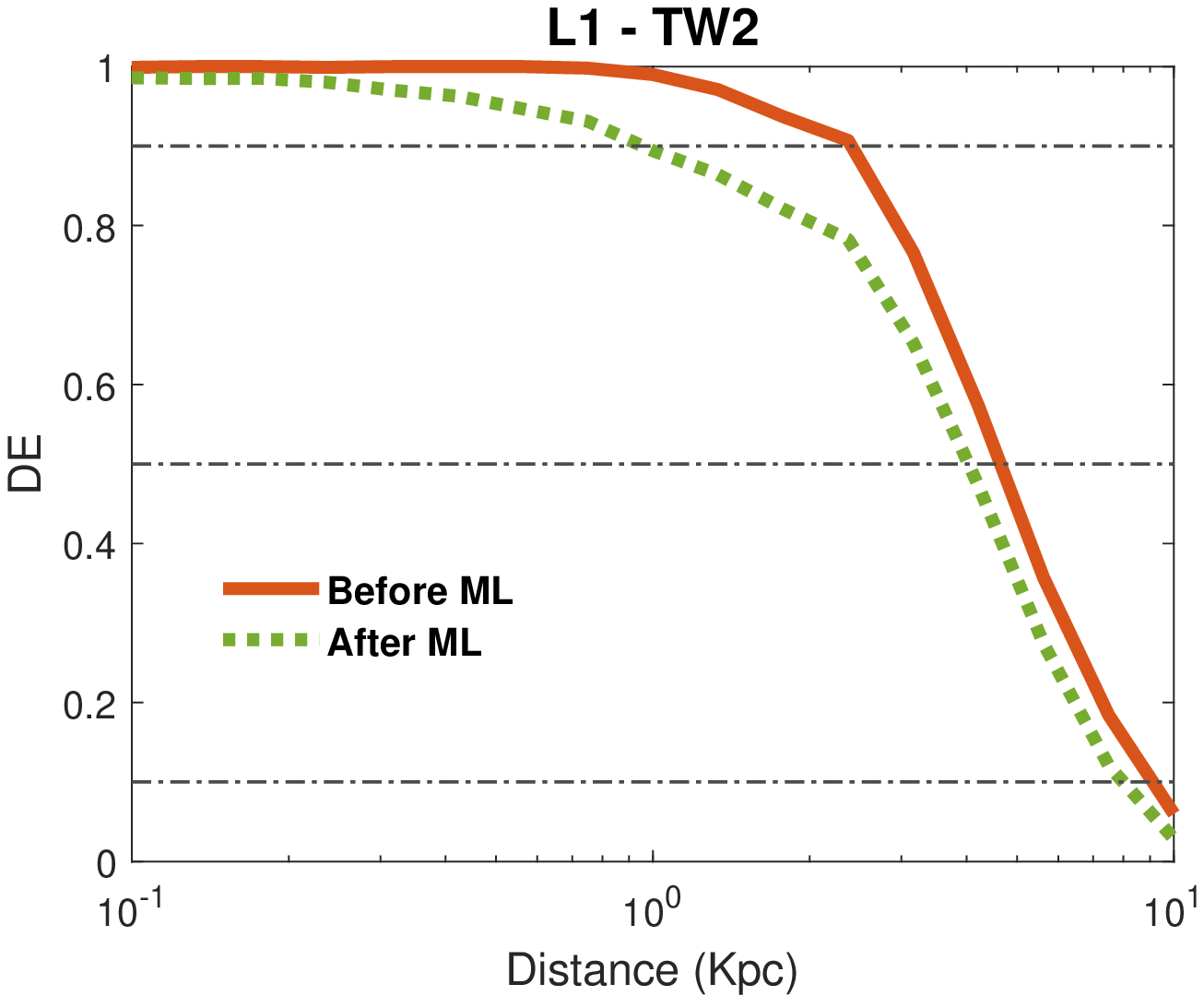}
    & 
    \includegraphics[width=0.26 \textwidth]{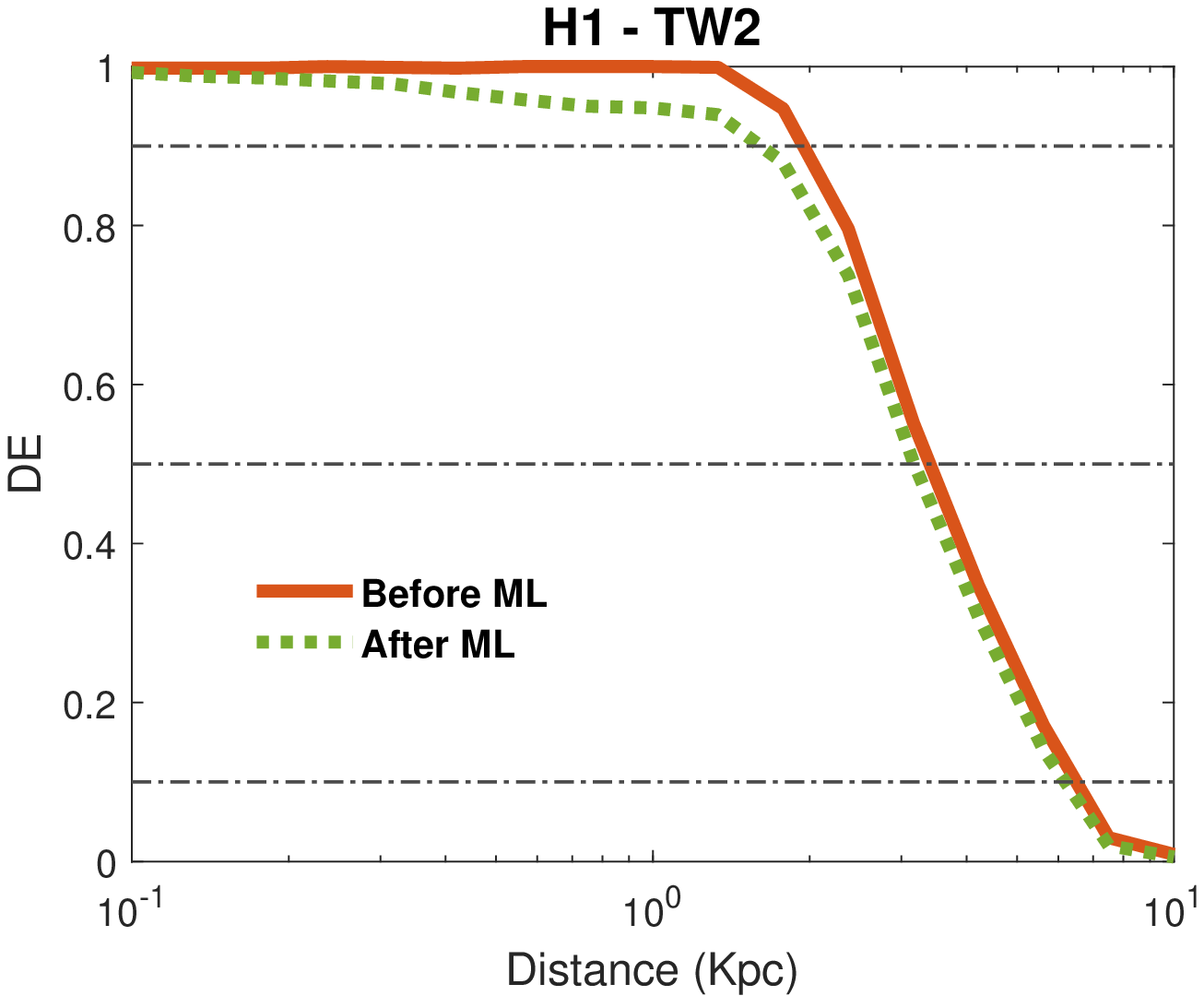}
    & 
    \includegraphics[width=0.26 \textwidth]{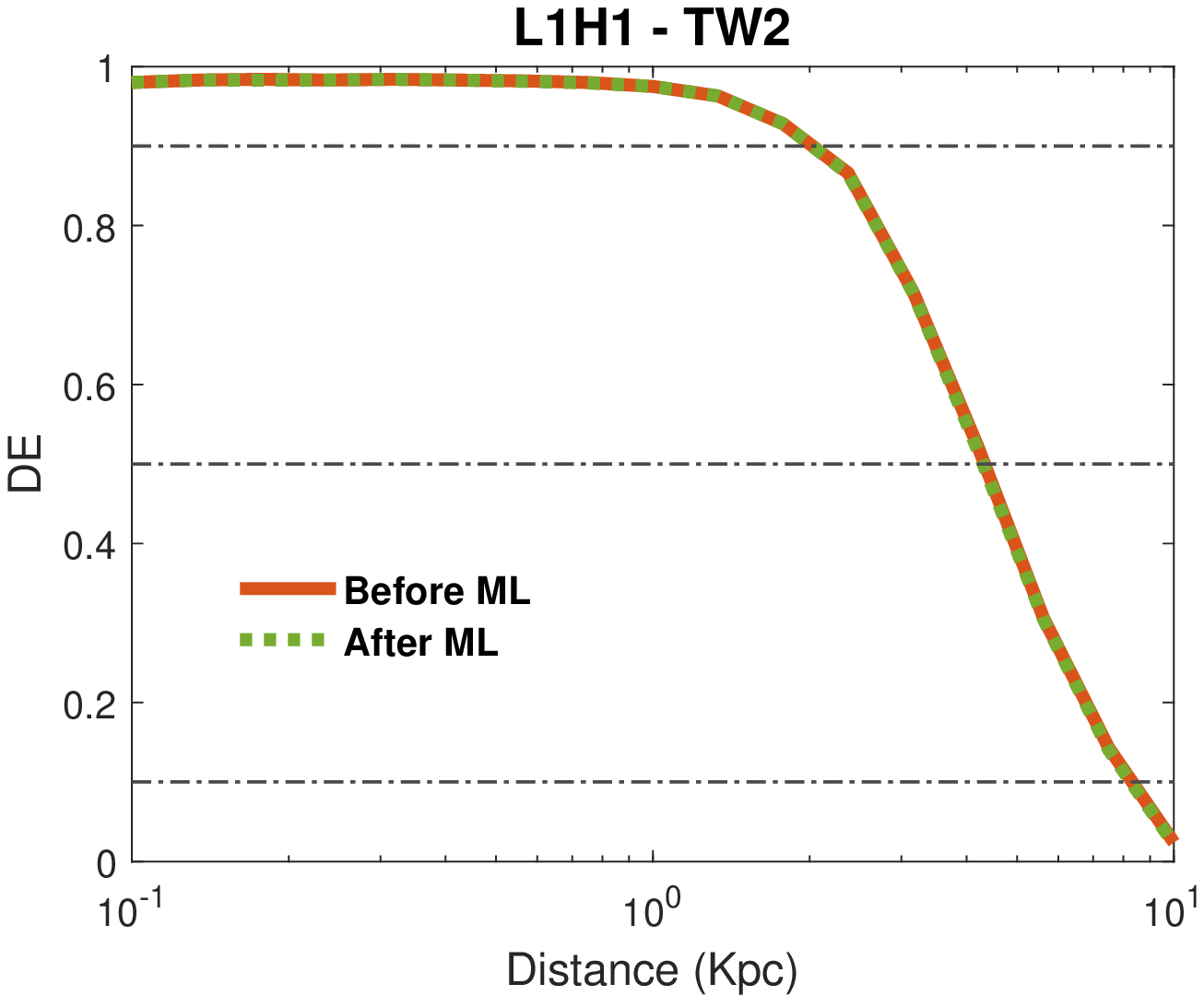}
    \\
    (a) & (b) & (c)
    \end{tabular}
    \caption{
    Results of the cWB simulation analysis for the time window TW2 before (red) and after (green) the application of the classification model.
    These results are the detection efficiency (DE) versus distance for each detector network, (a) L1, (b) H1, and (c) L1H1, obtained with the CCSNe GW he3.5 from \textit{Powell et al. 2019}.
    The FAR before/after the classification model is $2.97\times10^{-2}$/$1.78\times10^{-4}$, $2.53\times10^{-2}$/$1.95\times10^{-4}$, and $9.76\times10^{-6}$/$9.58\times10^{-9}$ for L1, H1, and L1H1, respectively.
    Again, for each detector network the classification model was trained with the corresponding dataset of triggers extracted from the time window DC1 and with signal triggers from all distances.
    }
    \label{fig:OS_DE_BeforeAfter}
\end{figure*}

Figure~\ref{fig:OS_DE_BeforeAfter} shows the DE versus distance before and after the application of the classification model for the CCSNe GW he3.5 from \textit{Powell et al. 2019} which enables to quantify the actual impact of the classifier.
There is a minimum signal reduction for single detector network and almost null effect for two detector network.
%
%
Horizontal lines allow to identify the distances with 90, 50 and 10\% of efficiency. For L1 and H1, there is no effect in the distance with 10\% of DE, while there is minimum reduction in the distance for 90 and 50\% DE. For L1H1, no effect in the distances is observed.
We conclude that for single detector networks there is marginal impact in the detection efficiency whereas there is no effect at all for the two detector network.
%
%

For the cWB analysis in time window TW2, table~\ref{Table:OS_NoiseAndSignalReduction} shows the noise and signal reduction whereas table~\ref{Table:OS_FARimprovementfactor} shows FAR improvement factor. 
%
%
As in the previous study, noise detection reduces and signal lost increases for larger distances and varies across the families of CCSNe GW.
Also, the two detector network provides the better performance with the higher noise reduction rates and lower signal reduction rates.
Equivalently, FAR improvement factor is higher for lower distances, and there are cases in which all noise triggers are completely identified and therefore the FAR after the classification is applied reaches its minimum value of the one over the background time.

\section{Conclusions}
\label{sec:conclusions}

This work investigated the use of supervised machine learning (ML) methods to recognize between noise and signal events using a set of reconstruction parameters from cWB outputs.
This constitutes a follow-up method devised to recognize and discard noise events while preserving signal events, which is essential to reduce the FAR and to increase the range of detection.

\begin{table*}
\caption{
Percentage of noise and signal reduction obtained by the three detector networks and for several distances of signal triggers. These results are for the time window TW2 and for each familiy of CCSNe GW waveforms. The classification model is trained from the dataset of noise and signal triggers obtained from the cWB analysis in the time window TW1. For the two detector network there are missing results because it was not possible to train and test the classification model since no signal triggers were obtained in the simulation analysis.}
\begin{tabular*}{\textwidth} {@{\extracolsep{\textwidth minus \textwidth}} c|cccccccccccc} 
\hline
\hline
\multicolumn{1}{c|}{ }
& \multicolumn{2}{c}{1kpc}
& \multicolumn{2}{c}{2.37kpc}
& \multicolumn{2}{c}{4.22kpc}
& \multicolumn{2}{c}{7.5kpc}
& \multicolumn{2}{c}{10kpc}
& \multicolumn{2}{c}{All} 
\\
 &Noise&Signal & Noise&Signal &Noise&Signal & Noise&Signal & Noise&Signal & Noise&Signal  \\
\hline
\hline
\multicolumn{1}{c}{  } & \multicolumn{12}{c}{ } \\
\multicolumn{1}{c}{L1} & \multicolumn{12}{c}{ } \\
\hline
%
%
%
Scheidegger et al. 2010 & $99.6$ & $4.9$ & $99.8$ & $8.4$ & $99.8$ & $12.7$ & $99.3$ & $16.4$ & $98.7$ & $17.7$ & $99.4$ & $8.3$ \\
O'connor at al. 2018    & $96.8$ & $19.7$ & $93.3$ & $31.7$ & $92.0$ & $35.7$ & $91.7$ & $35.3$ & $91.0$ & $35.2$ & $97.4$ & $17.8$ \\
Powell at al. 2019      & $99.3$ & $15.3$ & $98.3$ & $18.8$ & $97.1$ & $21.7$ & $93.6$ & $28.1$ & $91.9$ & $33.1$ & $99.4$ & $15.7$ \\
Powell at al. 2020      & $98.8$ & $8.7$ & $97.3$ & $12.5$ & $95.4$ & $15.4$ & $94.5$ & $19.2$ & $93.9$ & $21.2$ & $97.7$ & $11.2$ \\
Mezzacappa at al. 2020  & $99.2$ & $16.9$ & $96.8$ & $21.9$ & $94.2$ & $32.3$ & $93.9$ & $36.3$ & $93.2$ & $36.0$ & $99.1$ & $16.8$ \\
\hline
\multicolumn{1}{c}{  } & \multicolumn{12}{c}{ } \\
\multicolumn{1}{c}{H1} & \multicolumn{12}{c}{ } \\
\hline
%
%
%
Scheidegger at al. 2010 & $99.6$ & $4.8$ & $99.4$ & $8.2$ & $99.3$ & $11.9$ & $99.1$ & $15.9$ & $99.1$ & $16.8$ & $99.3$ & $8.0$ \\
O'connor at al. 2018    & $97.9$ & $22.4$ & $96.2$ & $39.2$ & $95.8$ & $40.2$ & $95.5$ & $40.4$ & $95.5$ & $40.2$ & $98.4$ & $17.8$ \\
Powell at al. 2019      & $99.7$ & $15.3$ & $98.7$ & $18.6$ & $97.8$ & $25.4$ & $95.9$ & $37.8$ & $95.3$ & $40.2$ & $99.2$ & $14.3$ \\
Powell at al. 2020      & $98.2$ & $6.9$ & $97.3$ & $12.1$ & $96.3$ & $16.3$ & $96.3$ & $20.4$ & $96.2$ & $22.2$ & $98.1$ & $11.6$ \\
Mezzacappa at al. 2020  & $99.4$ & $16.0$ & $97.6$ & $27.5$ & $95.4$ & $39.4$ & $95.2$ & $40.3$ & $95.5$ & $40.8$ & $99.1$ & $15.9$  \\
\hline
\multicolumn{1}{c}{  } & \multicolumn{12}{c}{ } \\
\multicolumn{1}{c}{L1H1} & \multicolumn{12}{c}{ } \\
\hline
%
%
%
Scheidegger at al. 2010 & $99.3$ & $1.0$ & $100.0$ & $1.5$ & $99.9$ & $0.8$ & $99.9$ & $0.0$ & $99.9$ & $0.1$ & $99.9$ & $1.7$ \\
O'connor at al. 2018    & $99.4$ & $0.8$ & - & - & - & - & - & - & - & - & $99.7$ & $0.7$ \\
Powell at al. 2019      & $100.0$ & $0.1$ & $99.9$ & $0.1$ & $99.4$ & $0.6$ & $98.2$ & $0.6$ & - & - & $100.0$ & $0.3$ \\
Powell at al. 2020      & $99.8$ & $0.1$ & $99.2$ & $0.5$ & $98.3$ & $2.8$ & $98.4$ & $6.0$ & $98.1$ & $5.9$ & $99.6$ & $1.9$ \\
Mezzacappa at al. 2020  & $100.0$ & $0.2$ & $99.7$ & $0.3$ & - & - & - & - & - & - & $100.0$ & $0.3$  \\
\hline
\hline
\end{tabular*}
\label{Table:OS_NoiseAndSignalReduction}
\end{table*}

\begin{table*}
\caption{
%
Improvement factor in the FAR (i.e., FAR before / FAR after) achieved for the cWB analysis in the time window TW2.
The classification model is trained from the dataset of noise and signal triggers obtained from the cWB analysis in the time window TW1.
Missing values indicate that no classification model was learned since no triggers were available. 
}
\begin{tabular*}{\textwidth} {@{\extracolsep{\textwidth minus \textwidth}} c|cccccc} 
\hline
\hline
\multicolumn{1}{c|}{ }
& \multicolumn{1}{c}{1kpc}
& \multicolumn{1}{c}{2.37kpc}
& \multicolumn{1}{c}{4.22kpc}
& \multicolumn{1}{c}{7.5kpc}
& \multicolumn{1}{c}{10kpc}
& \multicolumn{1}{c}{All} 
\\
\hline
\hline
\multicolumn{1}{c}{  } & \multicolumn{6}{c}{ } \\
\multicolumn{1}{c}{L1} & \multicolumn{6}{c}{ } \\
\hline
%
%
Scheidegger at al. 2010 & $260.7$ & $595.9$ & $417.1$ & $139.0$ & $74.5$ & $173.8$ \\
O'connor at al. 2018    & $30.9$ & $14.8$ & $12.6$ & $12.1$ & $11.1$ & $38.6$ \\
Powell at al. 2019      & $149.0$ & $57.9$ & $34.5$ & $15.7$ & $12.4$ & $166.8$ \\
Powell at al. 2020      & $80.2$ & $37.2$ & $22.0$ & $18.1$ & $16.4$ & $43.9$ \\
Mezzacappa at al. 2020  & $122.7$ & $31.4$ & $17.1$ & $16.4$ & $14.7$ & $115.9$  \\
\hline
\multicolumn{1}{c}{  } & \multicolumn{6}{c}{ } \\
\multicolumn{1}{c}{H1} & \multicolumn{6}{c}{ } \\
\hline
%
%
%
Scheidegger at al. 2010 & $242.7$ & $173.4$ & $140.0$ & $107.1$ & $117.5$ & $134.8$ \\
O'connor at al. 2018    & $48.5$ & $26.2$ & $23.6$ & $22.1$ & $22.1$ & $60.7$ \\
Powell at al. 2019      & $364.1$ & $77.5$ & $45.0$ & $24.6$ & $21.2$ & $130.0$ \\
Powell at al. 2020      & $55.2$ & $37.2$ & $27.4$ & $27.0$ & $26.6$ & $53.5$ \\
Mezzacappa at al. 2020  & $158.3$ & $41.4$ & $21.9$ & $20.7$ & $22.1$ & $113.8$  \\
\hline
\multicolumn{1}{c}{  } & \multicolumn{6}{c}{ } \\
\multicolumn{1}{c}{L1H1} & \multicolumn{6}{c}{ } \\
\hline
%
%
%
Scheidegger at al. 2010 & $145.6$ & $1019.0$ & $1019.0$ & $1019.0$ & $1019.0$ & $1019.0$ \\
O'connor at al. 2018    & $169.8$ & - & - & - & - & $339.7$ \\
Powell at al. 2019      & $1019.0$ & $1019.0$ & $169.8$ & $56.6$ & - & $1019.0$ \\
Powell at al. 2020      & $509.5$ & $127.4$ & $59.9$ & $63.7$ & $53.6$ & $254.8$ \\
Mezzacappa at al. 2020  & $1019.0$ & $339.7$ & - & - & - & $1019.0$  \\
\hline
\hline
\end{tabular*}
\label{Table:OS_FARimprovementfactor}
\end{table*}

%
The proposed follow-up ML method to enhance cWB searches was analyzed in two different studies.
The first aimed to ascertain the classification accuracy between noise and signal events (i.e., noise and signal reduction rates), while the second aimed to quantify the actual improvement in the statistical significance (i.e., reduction in FAR and FAP for the cases of on-source windows of 1 second and 1 day which represent a potential neutrino flux and optically targeted searches, respectively) and the impact in the detection efficiency.
The two studies considered different conditions as the number of detectors in the network (L1, H1, and L1H1); three stretches of open O3a train data; diverse distances of signal events; various CCSNe GW families with diverse characteristics; and three classification algorithms.
Overall, the results of these studies and variety of conditions showed high noise reduction rates greater than 90\% and low signal misclassification rates lower than 10\%.

Importantly, in our analyses we quantified the impact bring by the ML model in cWB offline searches.
The recognition and discharging of noise triggers from cWB outputs reduces the FAR in a factor of $\sim10$ for one detector (or even more for some families of GWs), and in a factor of $>100$ for two detectors in the network.
Notably, these FAR reductions are equivalent to an improvement in the statistical significance of $\sim 1.5 \sigma$ with one detector and even more with two detectors for the case of a on-source windows of 1 second.
For the case of optically targeted searches with on-source windows of 1 day, on the other hand, the FAR reduction yields to a $> 1 \sigma$ improvement with a two detector network.

%
The variety of explored conditions allowed to study several important aspects.
First, comparing the classification performance when tuning the ML model with a single or with various families of CCSNe GW showed no large differences in performance. This shows the robustness of the ML approach with respect to different types of CCSNe GW signatures, and points out that the morphology of noise and signal triggers are effectively distinguishable irrespective of the GW signals used to train and to test the classifier.vThis is noteworthy since in practice the ML model must be tuned with a pool of synthetic GWs obtained for example through numerical simulations (as is done in this work), but its usage in a real situation implies correctly identifying signal events generated by actual GWs that will not exactly match GWs used to train the model.
Secondly, there are high noise reduction rates with low signal reduction rates irrespective of the number of detectors in the network, though the classification performance is better with two detectors. This is simply because with two detectors the algorithm is tuned with a uniform distribution of noise triggers with low and moderate SNR since high SNR noise triggers do not survive coherence tests between detectors.
Also, regarding the different classification models, the results indicated a consistent superior performance with the support vector machine with radial basis function as kernel (SVMR). This model provides non-linear separation surfaces allowing to account for non-linearities in the feature space. Despite other methods can also be used for this problem, we recommend the use of SVMR since it has been proved to be stable and robust.

%
With regard to the training of the classification model, it is important to indicate that the training data in all analyses was always balanced, hence, there is an equal probability for each class (noise and signal). 
However, during the training procedure of the classifiers there is the possibility to assign more probability to one class. This would allow to increase the noise reduction at the expense of the signal reduction, or vice-verse.
This is an important aspect to be able to trade off between statistical significance and efficiency.
It is also possible to train the classification model using only a subset with loudest noise events, or with a subset of signal events with certain characteristics (duration, bandwidth, etc.).
This can be used to bias the follow-up ML model to prioritize specific aspects during the search by recognizing better one specific type of noise or signal event.

%
To sum up, this work presented a follow-up machine learning method for cWB based searches of GWs from CCSNe which can be use with one or multiple detectors.
The method identifies and discharges non-astrophysical noise transients allowing to reduce the false alarm rate, to improve the statistical significance and to increase the detection range of the searches.
The model is simply tuned with a set of noise and signal triggers and then can be easily incorporated into the search pipeline.

\section{Acknowledgements}

This research has made use of data, software and/or web tools obtained from the Gravitational Wave Open Science Center, a service of LIGO Laboratory, the LIGO Scientific Collaboration and the Virgo Collaboration.
We gratefully acknowledge the support of LIGO and Virgo for the provision of computational resources.
This material is based upon work supported by NSF’s LIGO Laboratory which is a major facility fully funded by the National Science Foundation.
Antelis and Moreno’s research is partially supported by CONACyT Ciencia de Frontera Project No. 376127.
Cavaglia's research is partially supported by NSF award number PHY 2011334 (2020).
Mukherjee’s research is partially supported by NSF award number PHY 1912630 (2019).


\bibliography{Paper.bib}

\end{document}